\documentclass[preprint, amsmath, amssymb, aps, pre]{revtex4-2}

\usepackage{dcolumn}
\usepackage{bm}
\usepackage{fancybox, graphicx}
\usepackage{subcaption}
\usepackage[dvipsnames]{xcolor}
\usepackage{multirow}
\usepackage{tikz}
\usetikzlibrary{matrix}
\usetikzlibrary{shapes.geometric, arrows, shadows}
\usetikzlibrary{fit,backgrounds} 

\newcommand{\wh}[1]{\widehat{#1}}
\newcommand{\wt}[1]{\widetilde{#1}}
\newcommand{\eps}{\epsilon}
\newcommand{\rmd}{\text{d}}

\newcommand{\Tm}[1]{\left< {#1} \right>} 
\newcommand{\Ttm}[1]{\langle {#1} \rangle} 
\newcommand{\Sm}[1]{\overline{#1}} 
\newcommand{\rms}[1]{{#1}_{\rm rms}} 

\newcommand{\abs}[1]{{\left|#1\right|}} 

\newcommand{\Sn}{\Phi} 
\newcommand{\snw}{\varphi} 

\newcommand{\td}{\tau_\rmd}
\newcommand{\tda}{\tau_{\rmd,\, \rm{wrong}}}
\newcommand{\tde}{\tau_{\rmd,\, \est}}
\newcommand{\tw}{\tau_{\rm w}}

\newcommand{\dt}{{\triangle_t}}

\newcommand{\Exp}{{\rm Exponential}}
\newcommand{\Ray}{{\rm Rayleigh}}
\newcommand{\Gam}{{\rm Gamma}}
\newcommand{\Par}{{\rm Pareto}}
\newcommand{\Unif}{{\mathcal{U}(0,2)}}
\newcommand{\Deg}{{\rm degenerate}}

\newcommand{\est}{{\rm est}}
\newcommand{\res}{{\rm res}}

\newcommand{\Eqref}[1]{Eq.~\eqref{#1}}

\newcommand{\Figref}[1]{Fig.~\ref{#1}}
\newcommand{\Figsref}[1]{Figs.~\ref{#1}}
\newcommand{\Secref}[1]{Sec.~\ref{#1}}
\newcommand{\Secsref}[1]{Secs.~\ref{#1}}
\newcommand{\Appref}[1]{App.~\ref{#1}}

\newcommand{\Tabref}[1]{Table~\ref{#1}}
\newcommand{\Tabsref}[1]{Tables~\ref{#1}}

\newcommand{\ldev}[1]{\textit{#1}}

\begin{document}

\title{Reconstruction of intermittent data time series as a superposition of pulses}

\author{Sajidah Ahmed}\email{sajidah.ahmed@uit.no}
\author{Odd Erik Garcia}\email{odd.erik.garcia@uit.no}
\author{Audun Theodorsen}\email{audun.theodorsen@uit.no}
\affiliation{%
Department of Physics and Technology, UiT The Arctic University of Norway, N-9037 Tromsø, Norway
}

\date{\today}

\begin{abstract}
Fluctuations in a vast range of physical systems can be described as a superposition of uncorrelated pulses with a fixed shape, a process commonly referred to as a (generalized) shot noise or a filtered Poisson process. In this contribution, we present a systematic study of a novel deconvolution method to estimate the arrival times and amplitudes of the pulses from realizations of such processes. The method shows that time-series can be reconstructed for various pulse amplitude and waiting time distributions. Despite a constraint on positive-definite amplitudes, it is shown that negative amplitudes may also be reconstructed by flipping the sign of the time series. The method performs well under moderate amounts of additive noise, both white noise and colored noise having the same correlation function as the process itself. The estimation of pulse shapes from the power spectrum is accurate except for excessively broad waiting time distributions. Although the method assumes constant pulse durations, it performs well under narrowly distributed pulse durations. The most important constraint on the reconstruction is information-loss, which limits the method to intermittent processes. The ratio between the sampling time and the average waiting time between pulses must be about 1/20 or smaller for a well sampled signal. Finally, given the system forcing, the average pulse function may be recovered. This recovery is only weakly constrained by the intermittency of the process.
\end{abstract}

\maketitle

\section{Introduction}\label{sec:introduction}

Intermittent and seemingly random fluctuations of order unity compared to the mean value are found in a variety of nonlinear physical systems, such as 
turbulence in neutral fluids \cite{salipante-2018, ding-2019, wang-2019, decristoforo-pof-2020} and atmospheric winds \cite{kristensen-1991, narasimha-2007}, water resources and hydrology
 \cite{claps-2005, lefebvre-2008},
complex fluids \cite{thomazo-2021},
fission chambers \cite{elter-2015}, physiology \cite{segal-1985,fesce-1986, richardson-2018} and biophysics \cite{hakim-2020} and
plasma turbulence, both simulations \cite{garcia-pop-2005-fluid,karimabadi-2013, anderson-2017, decristoforo-pop-2020}
and measurements from magnetically confined plasmas \cite{wootton-jnm-1990,antar-2003,dippolito-2011, garcia-nme-2017, kube-2020, militello-ppcf-2013, theodorsen-ppcf-2016, theodorsen-nf-2017, walkden-nf-2017}.
While such fluctuations may be extremely challenging to investigate from first principles based models, many fruitfully admit phenomenological modeling. One particularly useful reference model for time-series measurements from these systems is a stochastic model based on a superposition of localized pulses \cite{claps-2005, lefebvre-2008, elter-2015, theodorsen-ppcf-2016, richardson-2018, hakim-2020, thomazo-2021, garcia-nme-2017, decristoforo-pop-2020, decristoforo-pof-2020}. This model is called a shot noise process \cite{rice-1944, rice-1945} or a filtered Poisson process (FPP) \cite{parzen-sp}. If all pulses have the same functional shape and duration, the process can be written as a convolution between the pulse function (which may be considered a system response) and a random forcing. In the simplest case, the random forcing is driven by a Poisson process, giving the FPP its name. 

In many cases, it is of interest to extract amplitude and waiting-time statistics of the pulses or alternatively, the average pulse function if the forcing is known. One popular family of methods is conditional averaging \cite{pecseli-1989,oynes-1995,block-2006}, used, for example, in \cite{antar-2003,rudakov-2005,narasimha-2007,boedo-2014,garcia-nme-2017,kube-ppcf-2018,ding-2019}. Here, an amplitude threshold is set, and each time the signal crosses above this threshold, the time and amplitude of the peak are recorded along with the shape of the signal around the peak. Different authors use slightly different methods, and to the best of our knowledge, a systematic study of conditional averaging for the statistics of overlapping pulses in single-point time series is not available, although the case for non-overlapping structures has been investigated \cite{block-2006, teliban-2007} as has the case for two-dimensional structures with multi-point measurements \cite{johnsen-1987, pecseli-1989}. It is clear that both pulse overlap, and threshold requirement may influence the results and applicability of the conditional averaging method.

If the pulse function is known or may be estimated, some form of deconvolution may be performed to recover the forcing from realizations of the process. A method based on iterative deconvolution of the FPP has been shown to be robust in this problem \cite{theodorsen-pop-2018, kube-2020, decristoforo-pof-2020}. This algorithm, referred to as the Richardson-Lucy (RL) algorithm for Poissonian noise \cite{richardson-1972, lucy-1974} or the iterative space reconstruction algorithm (ISRA) for normally distributed noise \cite{benvenuto-2010} is not new; it was originally developed for image data in astronomy \cite{richardson-1972, lucy-1974} but has seen use in several other imaging applications \cite{dellacqua-2007, dellacqua-2010, dey-2006,benvenuto-2010}. This method requires a known or estimated common pulse function for all arrivals in the time series and reproduces the forcing in the maximum-likelihood sense under normally distributed and uncorrelated additive noise.  

In this contribution, we estimate pulse amplitudes and arrival times from the forcing by applying an iterative deconvolution on realizations of the process. Our main aim is to report accurately on the prospects and limitations of using this procedure as a tool for time series analysis. A wide variety of assumptions regarding pulse overlap, pulse functions, amplitude and waiting time distributions, additive noise, and correlations between these may be made in different contexts and applications. Further, noise, pulse superposition, and distribution of pulse shapes put limits on the estimation of pulse amplitudes and arrivals. To limit the scope of this contribution, we will focus on assumptions relevant for time series measurements in turbulent fluids and plasmas, and in particular at the boundary of magnetically confined fusion plasmas which are characterized by broad and positive definite amplitude distributions, close to Poisson distributed arrivals and asymmetric, positive and exponentially decaying pulses \cite{antar-2003,dippolito-2011,militello-ppcf-2013,theodorsen-ppcf-2016, walkden-nf-2017, theodorsen-nf-2017, garcia-nme-2017, kube-2020, theodorsen-pop-2018, rudakov-2005,boedo-2014}.

We note that the deconvolution algorithm may also be used the other way around: If the forcing is known, the pulse function may be estimated using the same algorithm. In this way, it may be used to find the system response to a controlled input forcing, if the assumption of linearity is satisfied.

This contribution is structured as follows: In \Secref{sec:theory} we review the stochastic model, define the base case of the model which will be modified during the following investigations, and present the deconvolution algorithm. In \Secref{sec:limitations-overlap} the limitations of the signal reconstruction from estimated amplitudes and arrival times due to sampling and pulse overlap are investigated. In \Secref{sec:rec-amp-wait} we assess the ability of the algorithm to reproduce various amplitude and arrival times for a known pulse function. Then, in \Secref{sec:noise} distortions due to additive noise, both uncorrelated and correlated, are investigated and a criterion for noise removal is established. Following this, in \Secref{sec:td-est} we report on the effect of estimating the pulse shape assuming a known functional form. Both the effects of over- and underestimating the pulse duration are considered, as well as the effect of a narrow distribution of pulse durations. Lastly, we turn the question around in \Secref{sec:find-pulse-shape} and look at how well an unknown pulse function can be estimated if the pulse arrivals and amplitudes are known. Finally, we discuss the results and conclude in \Secref{sec:discussion-conclusion}. In this contribution, we will not consider the effects of statistical convergence: we will always use long time series to avoid large uncertainty or bias in parameter estimates and will always run the deconvolution algorithm to convergence. Effects of short time series on moment estimation were previously investigated in Refs.~\onlinecite{kube-2015,garcia-pop-2016}. \Secsref{sec:limitations-overlap}--\ref{sec:find-pulse-shape} are largely independent; the reader may consult to the problem of particular interest. The numerical implementation of this method with its library of functions is openly available on the GitHub page of the UiT Complex Systems Modelling group \cite{csm-github}.

\section{Theory}\label{sec:theory}
In this section, we first review the stochastic model given by a superposition of pulses with fixed shape and duration. This is followed by a presentation of the deconvolution algorithm. The section concludes by discussing how pulse amplitudes and arrival times can be recovered from realizations of the stochastic process using the deconvolution algorithm. Throughout this manuscript, we adopt the notation that angular brackets $\Ttm{\cdot}$ refers to a theoretical mean value, while an overline $\Sm{\cdot}$ refers to a sample mean. Quantities estimated by the deconvolution method use the est-subscript $\cdot_\est$.

\subsection{The stochastic model}

The basic stochastic model considered here is a superposition of pulses with a fixed shape defined as
\begin{equation} \label{eq:fpp-def}
    \Sn_K(t) = \sum\limits_{k=1}^{K(T)} A_k \snw\left(\frac{t-s_k}{\tau_k} \right).
\end{equation}
Here, $K(T)$ denotes a point process on the interval $[0,T)$ with sorted event arrival times $s_k$ and waiting times $w_k = s_{k}-s_{k-1}$ with mean value $\Ttm{w}$. We will in general restrict $K(T)$ to be a \emph{renewal} process, where the waiting times are independently and identically distributed \cite{parzen-sp}. The amplitudes $A_k$ are randomly distributed with mean value $\Ttm{A}$, and in general assumed to be positive definite. The pulse function $\snw$ is assumed to be the same for all events but may have randomly distributed duration times $\tau_k$. The average pulse duration time is denoted by $\td = \Ttm{\tau}$. In the following, we take all $A_k$, $w_k$ and $\tau_k$ to be independent random variables, and each variable family is independently and identically distributed. 

The fundamental parameter of the stochastic model is $\gamma = \td/\Ttm{w}$, referred to as the \emph{intermittency parameter}. It describes the degree of pulse overlap and quantifies how intermittent the fluctuations are through the skewness and flatness moments \cite{garcia-prl-2012}. In general, the influence of $\gamma$ on the qualitative appearance and features of the process is the following: For $\gamma$ of order unity or smaller, pulses appear well separated due to the waiting times between pulses being long compared to the duration time. When $\gamma < 1$, the process $\Sn$ will thus have a small mean value compared to the mean pulse amplitude and large relative fluctuation levels. For $\gamma$ much larger than one, there is significant pulse overlap due to short waiting times and long pulse durations. This results is a process that is near normal distributed, where $\Sn$ will have a large mean value and small relative fluctuation levels \cite{garcia-prl-2012}. In \Secref{sec:limitations-overlap}, we will see that it is not the intermittency parameter which determines our ability to estimate the amplitudes and arrival times of the process, but the ratio between the sampling time and the average waiting time. We write this ratio as $\gamma \theta$, where $\theta=\triangle_t/\tau_\text{d}$ is the sampling time $\triangle_t$ normalized by the average pulse duration.  

Here, some care must be taken regarding the sampling process. If we consider a sampling method which instantaneously measures a value at regular intervals, pulses which are only a few sampling times in duration may be undersampled, such that the true amplitude of the pulse is not captured. Thus the process may contain an artificially low number of pulses or have an artificially low mean value compared to an adequately sampled process. If the sampling time is an \emph{integration time} such as a camera exposure time, all pulses will be identified. However, there may be distortions in the pulse shape due to the integration, as the integration smears out the pulse function. In either case, the deconvolution method described in \Secref{sec:deconv} reproduces the \emph{sampled} time series and is not designed to make inferences regarding missing or distorted pulses. Thus, an accurate reconstruction of the true process also depends on a sufficiently high sampling rate to resolve the pulses. Our testing suggests that for the exponential pulse function in \Eqref{eq:def-exp-one} (below), a sampling time of $10$ times the typical pulse duration time or better is sufficient, that is, $\triangle_t/\tau_\text{d}\leq10^{-1}$.

We take the following as the \emph{base case}:
\begin{itemize}
    \item $K(T)$ is a Poisson process. Therefore, it follows that the arrival times are independent and uniformly distributed on the interval $[0,T)$, and the waiting times $w_k = s_{k}-s_{k-1}$ are independently and exponentially distributed with mean value $\Ttm{w}$ \cite{parzen-sp}. The mean value of the Poisson process is then given by $\Ttm{K(T)} = T/\Ttm{w}$.
    
    \item Degenerate distribution of pulse durations, $\td P_\tau(\tau)=\delta(\tau-\td)$, so that all pulses have the same duration, $\tau_k=\td$.
    
    \item Fixed one-sided exponential pulse function for all events given by a jump followed by an exponential decay,
\begin{equation}  \label{eq:def-exp-one}
  \snw\left(x\right)=\begin{cases}  
      0, & x<0,\\
      \exp(-x), & x \geq 0. \end{cases}     
\end{equation}.

\item Exponentially distributed amplitudes with mean value $\Ttm{A}$, $\Ttm{A} P_A(A)=\exp{(-A/\Ttm{A})}$ for $A>0$.
\end{itemize}
Some consequences of this base case are of interest \cite{garcia-prl-2012,theodorsen-ps-2017,garcia-pop-2017-2,theodorsen-ppcf-2018}:
\begin{itemize}
    \item The probability distribution function (PDF) of $\Sn$ is a Gamma distribution with shape parameter $\gamma = \td/\Ttm{w}$ and scale parameter $\Ttm{A}$,
    \begin{equation}\label{eq:pdf-base-sn}
    P_\Sn(\Phi; \gamma, \Tm{A}) = \frac{\Phi^{\gamma-1}}{\Tm{A}^\gamma \Gamma(\gamma)} \exp\left( -\frac{\Phi}{\Tm{A}} \right), \quad \Phi>0.
    \end{equation}
    
    \item The four lowest order moments are the mean $\Ttm{\Sn} = \gamma \Ttm{A}$, variance $\rms{\Sn}^2 = \gamma \Ttm{A}^2$, skewness $S_{\Sn} = \Ttm{(\Sn - \Ttm{\Sn})^3}/\rms{\Sn}^3 = 2/ \sqrt{\gamma}$ and flatness (or excess kurtosis) $F_{\Sn} =  \Ttm{(\Sn - \Ttm{\Sn})^4}/\rms{\Sn}^4 - 3 = K_\Sn -3 = 6/\gamma$, where $K_\Sn$ is the kurtosis. 
    
    \item The frequency power spectral density (PSD) of $\Sn$ has a Lorentzian shape, 
    \begin{equation}\label{eq:psd-base-sn}
    \Omega_\Sn(\omega) = \rms{\Sn}^2 \frac{2 \td}{1 + \td^2 \omega^2} + 2\pi \Tm{\Sn}^2 \delta(\omega).
    \end{equation}

\end{itemize}

In the base case, there are three fundamental model parameters: $\gamma$, $\Ttm{A}$ and $\td$. From realizations of the process, the first two may be estimated from the PDF given by \Eqref{eq:pdf-base-sn}. From these follows the mean value and the standard deviation of the process. The final parameter $\td$ may be estimated from the auto-correlation function or from the frequency PSD given by \Eqref{eq:psd-base-sn}. 

In the case that all pulses have the same duration $\tau_\text{d}$, we may express the stochastic model as a convolution between the pulse function $\snw$ and a forcing $f_K$,
\begin{equation} \label{eq:sn-conv}
    \Sn_K\left(t\right) = \left[\snw * f_K\right]\left(\frac{t}{\tau_\mathrm{d}}\right),
\end{equation}
where $f_K$ is given by
\begin{equation} \label{eq:def-fk}
    f_{K}\left(t\right) = \sum\limits_{k=1}^{K\left(T\right)} A_k \delta\left(\frac{t-s_k}{\tau_\mathrm{d}}\right).
\end{equation}
Hence, one can say that $\Sn_K$ is a train of delta pulses, given by $f_K$, arriving according to the point process $K(T)$ which is passed through a filter $\snw$. The FPP may be considered a linear model for highly nonlinear phenomena, where the nonlinearity has been baked into the distributions of $A$ and $s$ and the pulse function $\snw$. Given an estimate of the pulse function, it is therefore possible to estimate $f_K$ by deconvolving a realization $\Sn_K$ with the pulse function $\snw$. Further discussions on estimation of model parameters from realizations of the process are given in Refs.~\onlinecite{kube-2015,garcia-pop-2016,theodorsen-ps-2017,theodorsen-pop-2018}.

\subsection{Deconvolution method}\label{sec:deconv}

To estimate the pulse arrival times $s_k$ and amplitudes $A_k$ from realizations of the stochastic process, the ``iterative image space reconstruction algorithm'' (ISRA), is presented in Ref.~\onlinecite{benvenuto-2010}, and we point the reader to this publication for a more thorough discussion of the details of the algorithm. Here, we consider $\Sn$, $\snw$ and $f$ to be discretized with a uniform sampling time $\dt$ and an odd number of data points $N$.  Furthermore, $\Sn$ is assumed to be corrupted by uncorrelated white noise denoted $X$, so that we may write for $0 \leq j < N$ 
\begin{equation} \label{eq:sn-conv-disc}
    \Sn_j = (\snw * f)_j + X_j
\end{equation}
where $f_j$ is a sum of Kronecker deltas weighted by the pulse amplitudes, $f_j = \sum_{k=1}^{K\left(T\right)} A_k  \delta_{j,j_k}$, and by abuse of notation $\snw_j = \snw(j \dt/\td) $.
We have suppressed the subscript $K$ on $\Sn$ and $f$ for simplicity of notation and $j_k$ is $s_k/\dt$ rounded to the nearest integer. In the following, the hat symbol $\wh{\cdot}$ is used to denote a flipped vector, $\widehat{\snw}_j = \snw_{N-1-j}$. 

The maximum-likelihood estimation applied to the above model leads to the optimization problem 
\begin{eqnarray}\label{eq:deconv-opt-prob}
    {\rm minimize\,}& J_\Sn(f) = \frac{1}{2} \| \snw * f - \Sn \|^2 \\
    {\rm subject \: to}&\, f \geq 0.
\end{eqnarray}
The iteration scheme
\begin{equation} \label{eq:deconv-scheme}
    f^{(n+1)}_{j} = f^{(n)}_{j}\frac{\left(\Phi * \widehat{\varphi}\right)_j + b}{\left(f^{(n)} * \varphi * \widehat{\varphi}\right)_j + b}
\end{equation}
is known to converge asymptotically to the least-squares solution of the optimization problem under certain conditions on $\snw$ \cite{benvenuto-2010}. For our purposes, $\snw(t) \geq 0$ and $\snw(t=0) >0$ are sufficient conditions. Here, $b$ is a free parameter chosen such that $\left(\Sn * \widehat{\snw}\right)_j + b > 0\,\forall\,j$. 
The method is designed to extract a \emph{non-negative} signal $f$ from a measurement described by \Eqref{eq:sn-conv-disc}, where the \emph{only} negative parts of the signal are due to noise. The effect of negative pulse amplitudes is explored in \Secref{sec:rec-amp-wait}. We note that the standard deviation of the noise $X$, or alternatively the signal-to-noise ratio, plays no role in the iteration scheme.

Numerical testing reveals that the choice of the initial guess $f^{(0)}$ as well as the exact value of $b$ may play a role in the rate of convergence but do not affect the result of the iteration given by \Eqref{eq:deconv-scheme} if $b$ is small compared to the mean signal value. Consequently, we set the initial guess to a positive constant and the $b$-parameter to
$b=10^{-10}-b_{\rm min}$, where $b_{\rm min} =\min[0, \min{(\Sn * \hat{\snw})}]$. The small constant $10^{-10}$ is added to avoid issues with the division of numbers close to zero in the denominator in \Eqref{eq:deconv-scheme}.

\subsection{Extracting amplitudes and arrival times}\label{sec:extract-amp-arrivals}

\begin{figure}
    \centering
    \includegraphics[width=0.9\textwidth]{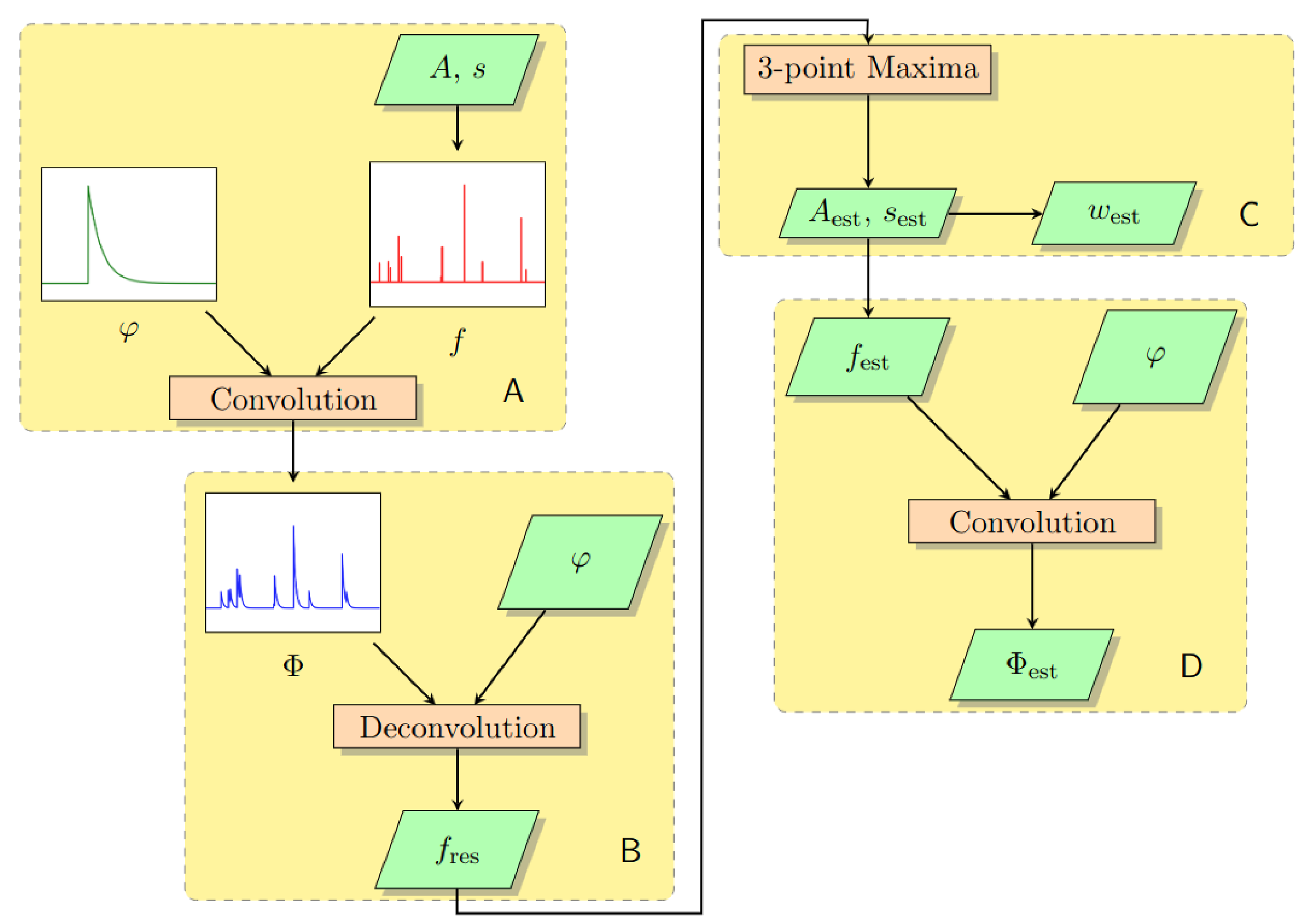}
    \caption{Flow chart to show the procedure applied in this study of the deconvolution algorithm. The flow chart is detailed in the main text. For experimental data, the processes in boxes $\mathsf{B}$, $\mathsf{C}$ and $\mathsf{D}$ would be applied. In this contribution, the focus is on the effect of $\mathsf{C}$ and $\mathsf{D}$.\label{fig:flow-chart}}
\end{figure}

The result of the deconvolution is the maximum-likelihood estimate of the forcing, denoted $f_\res$ where the subscript $\cdot_\res$ stands for `result'. In this study, we are interested in estimating the amplitudes and arrival times that define the forcing $f$. For the realizations investigated in this contribution, the delta-train forcing is estimated as a series of sharply localized peaks. Therefore, a peak-finding algorithm must be applied to recover the pulse arrival times and amplitudes. We have chosen to employ a simple 3-point running maxima, so that a data point $f_{\res, j}$ is classified as a peak if it is larger than both neighboring points: $f_{\res, j}>f_{\res, j\pm1}$ for all $1\leq j < N-1$. The estimated arrival time $s_\est$ is the location of the maxima and the estimated amplitude $A_\est$ is the value of the maxima. 

Using the found maxima directly leads to far too many detected events for small ($\gamma<1$) intermittency parameters (6 times the true number of events for $\gamma = 1/10$). This is due to numerical noise in sections of $f_\res$ without true pulses. To remove these spurious pulses, we introduce a small amplitude threshold equal to $10^{-2} \Ttm{A}$, which corresponds to the size of $\abs{f_\res - f}$ for $\gamma = 1/10$. The presence of this threshold does not influence the results in the absence of additive noise, as investigated below. In the presence of additive noise, a stricter amplitude threshold is placed on local maxima. This amplitude threshold is discussed further in \Secref{sec:noise}.

The entire reconstruction process is described by the flow diagram in \Figref{fig:flow-chart}. In box $\mathsf{A}$, the synthetic signal $\Sn$ is generated. The waiting times $w$ are randomly drawn until their sum exceeds the specified signal duration $T$. The arrival times $s$ are computed from the waiting times, and a number of amplitudes $A$ matching the arrival times are randomly drawn. The synthetically generated forcing $f$ consists of the amplitudes $A$ and the arrival times $s$, and convolving this forcing with the specified pulse function $\varphi$ gives the synthetic signal $\Phi$. In box $\mathsf{B}$, we perform the ISRA specified in \Eqref{eq:deconv-scheme}, where $f_\res$ is the result of the deconvolution. Here, it may be necessary to estimate $\snw$, for example from the PSD as discussed in \Secref{sec:td-est}. In box $\mathsf{C}$, $A_\est$, $s_\est$ and $w_\est$ refer to the amplitudes, arrival times and waiting times extracted using the 3-point maxima method described above. Finally, in box $\mathsf{D}$, the signal is reconstructed. Here, $f_\est$ refers to forcing which consists of $A_\est$ and $s_\est$. $\Sn_\est$ refers to the signal reconstructed by using the estimated amplitudes and arrival times. For the model realizations considered in this contribution, it will be shown that the process in box $\mathsf{B}$ in \Figref{fig:flow-chart} captures $f$ very well. Thus, the focus of this contribution is the prospects and limitations of $\mathsf{C}$ and $\mathsf{D}$.

To check that the mean values of the estimated variables have converged, we employ a bootstrapping technique \cite{Davison_Bootstrap_1997}. For a given data set $X$ containing $N$ samples, we first estimate the sample mean $\Sm{X}$. Then we draw $3N/4$ random samples with replacement (that is, we are allowed to draw the same sample multiple times) and estimate the sample mean $\Sm{X}_{3N/4}$. Repeating this procedure 100 times gives a data set containing 100 samples of $\Sm{X}_{3N/4}$. Estimating the standard deviation of this data set gives a measure of how well $\Sm{X}$ has converged. A large standard deviation would indicate that we have too few samples $N$, and should repeat our estimate with a larger data set. We chose $3N/4$ samples instead of $N$ for the random draw to err on the side of overestimating the standard deviation. Estimates for higher moments proceed analogously. In all cases, this produced standard deviations within $6\%$ of the corresponding mean values, indicating well-converged statistics.

Since the parameter values (in particular $\gamma$) are varied over an order of magnitude, we report the bias in the estimate as the \emph{ratio} between the estimated value and the true value instead of the more common \emph{difference} between the estimated value and the true value. This lets us compare estimates for different input parameters directly. In this contribution, a deviation of 10\% or more from the true value is considered significant and is indicated with \textit{italic} numbers in the tables. This will be the basis for our discussion of the results.

\section{Estimating amplitudes and waiting times}\label{sec:limitations-overlap}

Two effects lead to loss of information of the true pulse arrivals in realizations of the process: (1) The point process $K(T)$ is continuous but as we are working with discrete time series, pulses closer than a sampling time cannot be separated.  These will be counted as one pulse arrival with amplitude equal to the sum of their amplitudes, corrupting the resulting estimated amplitude distribution and number of pulses. (2) As the peaks of $f_\est$ may have finite widths, a peak-finding algorithm must be employed. A straightforward and permissive method, a 3-point maxima, compares each data point to its neighboring points and flags it as a peak if it is bigger than both neighbors. This method retains only the highest amplitude event if two or more pulses arrive closer than two sampling times to each other.

In \Appref{app:number-events}, it is shown that in the base case, the average number of 3-point maxima $M$ in $f$ as compared to the average number of pulses $\Ttm{K}$ is given approximately by
\begin{equation}\label{eq:approx-mean-M-vs-K}
    \frac{\Ttm{M}}{\Ttm{K}} \approx \frac{1 - \exp{(-3 \gamma \theta)}}{3 \gamma \theta}.
\end{equation}
Here, $\theta = \dt/\td$ is the normalized sampling time and $\gamma \theta = \dt/\Ttm{w}$. In order to recover approximately 90\% of the pulses using the 3-point maxima, $\gamma \theta < 0.075$ is necessary, while to recover approximately 95\% of the pulses, $\gamma \theta <0.035$ is necessary. This suggests an approximate threshold of $\gamma \theta \leq 1/20$ to recover most (above 90\%) of the pulses.  In this contribution, we have set $\dt = 10^{-2} \td$, leading to an assumed approximate condition $\gamma<5$ for reconstruction of amplitude and arrival time distributions within 10\% variation in average values. We will see that in general, $\gamma = 10$ gives too much overlap while in some cases, $\gamma \leq 1$ is required, which corresponds to approximately 98.5\% pulse recovery. This threshold should be seen as a guideline, dependent on the precision and objectives required for application of the method.

\begin{figure}
    \centering
    \includegraphics[width=1.\textwidth]{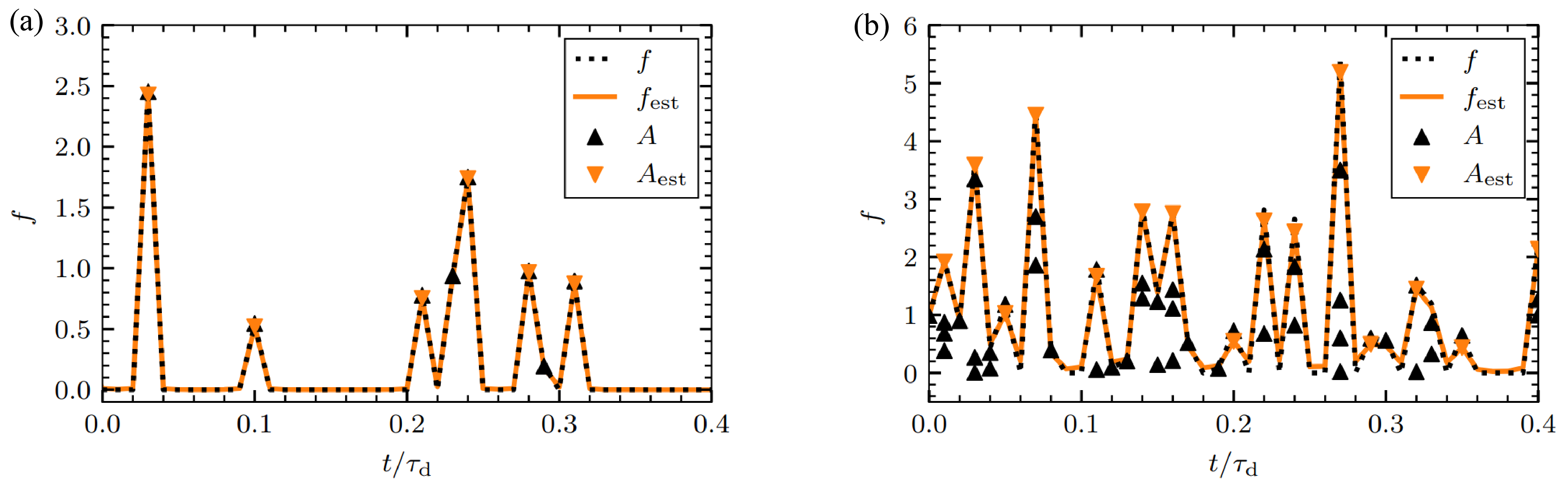}
    \caption{Comparison between original forcing $f$ and the estimated forcing $f_\mathrm{est}$ for two different intermittency parameters, (a) $\gamma = 10$ and (b) $\gamma=10^2$, no noise $X=0$, exponentially distributed pulse amplitudes and a normalized sampling time of $\dt/\td=10^{-2}$. The exponential pulses are uncorrelated and have a uniform distribution of arrival times. The markers denote the original amplitudes $A$ and the estimated amplitudes $A_\mathrm{est}$.}
    \label{fig:diff-gamma-res}
\end{figure}

The sampling and local maxima issues are illustrated in \Figref{fig:diff-gamma-res}, which shows results from deconvolution for fixed $\theta$ and two different values of $\gamma$. Although the estimated forcing (the orange solid line) is a good approximation of the original forcing signal (the black dotted line), even for $\gamma = 10^{2}$, the same cannot be said of the estimated amplitudes (the orange triangles) when compared to the original amplitudes (the black triangles). In the case of $\gamma=10$, the error is moderate and mainly caused by the 3-point maxima not identifying arrivals at neighboring points. The effect is more severe in the case $\gamma = 10^{2}$, where it is evident that $f_\text{est}$ is given by a superposition of multiple arrivals closer than one time step of each other. 

\begin{figure}
    \centering
        \includegraphics[width=1.\textwidth]{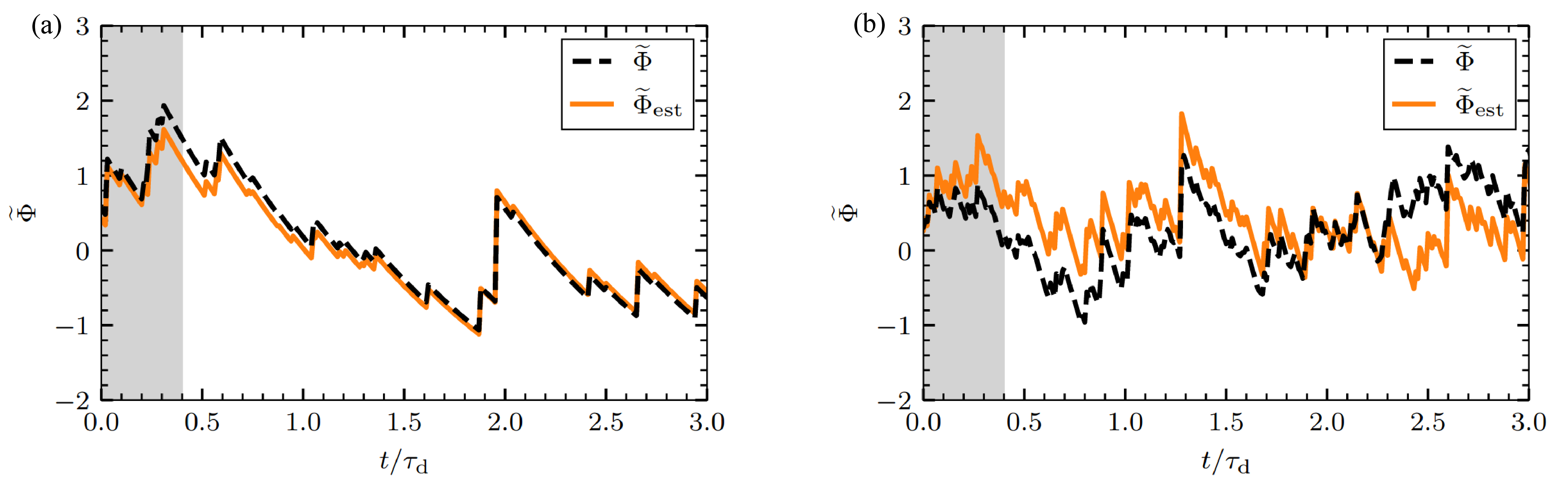}
    \caption{Comparison between normalized original time series $\widetilde{\Phi}(t)$ and the normalized reconstructed time series $\widetilde{\Phi}_\mathrm{est}(t)$ for two different intermittency parameters, (a) $\gamma = 10$ and (b) $\gamma=10^2$, using amplitudes and arrival times estimated with the deconvolution method. The forcing for the first $0.4$ normalized time units is presented in \Figref{fig:diff-gamma-res} and is gray shaded here.}
    \label{fig:diff-gamma-sig-rec}
\end{figure}

The corresponding original and reconstructed time series are presented in \Figref{fig:diff-gamma-sig-rec} for one-sided exponential pulses and an exponential pulse amplitude distribution. 
Here and in the following, $\wt{\Sn}$ denotes the normalized variable  $\wt{\Sn} = \left(\Sn-\Ttm{\Sn}\right)/\rms{\Sn}$.
 For these high $\gamma$-values, the FPP resembles a normally distributed process. The reconstructed signal largely follows the overall path of the original signal but deviates in detail due to finite sampling rate and pulse overlap as described above. These differences may have a profound influence on the estimation of the amplitude and waiting time statistics, as investigated in detail below.

In \Tabref{tab:est-moments-various-gamma} the first four moments as estimated from the reconstructed time series are presented for various values of the intermittency parameter. The estimated moments are normalized by the sample moments of the original time series. In all cases, one-sided exponential pulses with an exponential amplitude distribution are used. For $\gamma \leq 5$, there is at most 6\% disagreement between the estimated and sample moments, within our allowed tolerance. For $\gamma \geq 10$, both the mean value and standard deviation are underestimated, with values below the 10\% threshold. The underestimation is worse for larger intermittency parameters. Higher-order moments are wrongly estimated for $\gamma \geq 50$, as evidenced by the three values in the lower right-hand corner of the table. Thus, we conclude that \Tabref{tab:est-moments-various-gamma} provides evidence for the condition $\gamma \theta \leq 1/20$ for accurate reconstruction.

The deviation in the first two moments may be explained by the following: Due to pulse overlap and the 3-point maxima, several true events are discounted. This leads to an underestimation of the mean value and the standard deviation, more severely for higher intermittency parameters. As the discounted events are preferentially small, the small-amplitude variations are decreased, leading to a lower overall rms-level of the signal. By taking the square root of the second column and dividing it by the first column, it is seen that the relative fluctuation level, however, is robust. The deviations in the higher moments are less systematic, and a Monte-Carlo study should be carried out to put reasonable error bars on these values. The important conclusion from this table is that it supports the approximate condition $\gamma \theta \leq 1/20$ and therefore a time-consuming Monte-Carlo study for the largest intermittency parameters has not been performed here.

\begin{table}
\begin{ruledtabular}
\centering 
 \begin{tabular}{c||c c c c} 
\multirow{2}{*}{$\gamma$} & \multicolumn{4}{c}{Moments} \\
   & $\Sm{\Sn}_{\rm{est}}/\Sm{\Sn}$ & $\Sn_{\rm{rms,est}}^{2}/\Sn_{\rm{rms}}^{2}$ & $S_{\rm{\Sn,est}}/S_\Sn$ & $(F_{\rm{\Sn,est}}-3)/(F_\Sn-3)$\\ [0.5ex] 
 \hline 
 $10^{-1}$ & 1.00 & 0.99 & 1.00 & 1.00\\
 1         & 0.99 & 0.97 & 0.99  & 0.99 \\  
 5         & 0.97 & 0.94  & 0.99 & 0.97 \\ 
 10        & 0.94 & \ldev{0.78}  & 0.98 & 0.94 \\ 
 50        & \ldev{0.78}  & \ldev{0.60} & 1.02 & \ldev{0.58} \\ 
 $10^{2}$  & \ldev{0.68}  & \ldev{0.50} & \ldev{1.13} & \ldev{1.32} 
 \end{tabular}
\caption{Table showing the sample moments of the reconstructed time series for various intermittency parameters $\gamma$ as compared to the sample moments of the model realization. $10^{5}$ iterations were used for parameters $\theta = 10^{-2}$, $T/\td = 10^{4}$ and no noise was added.  The theoretical expressions for the moments are given directly below \Eqref{eq:pdf-base-sn}.}
\label{tab:est-moments-various-gamma}
\end{ruledtabular}
\end{table}

\begin{figure}
    \centering
    \includegraphics[width=1.\textwidth]{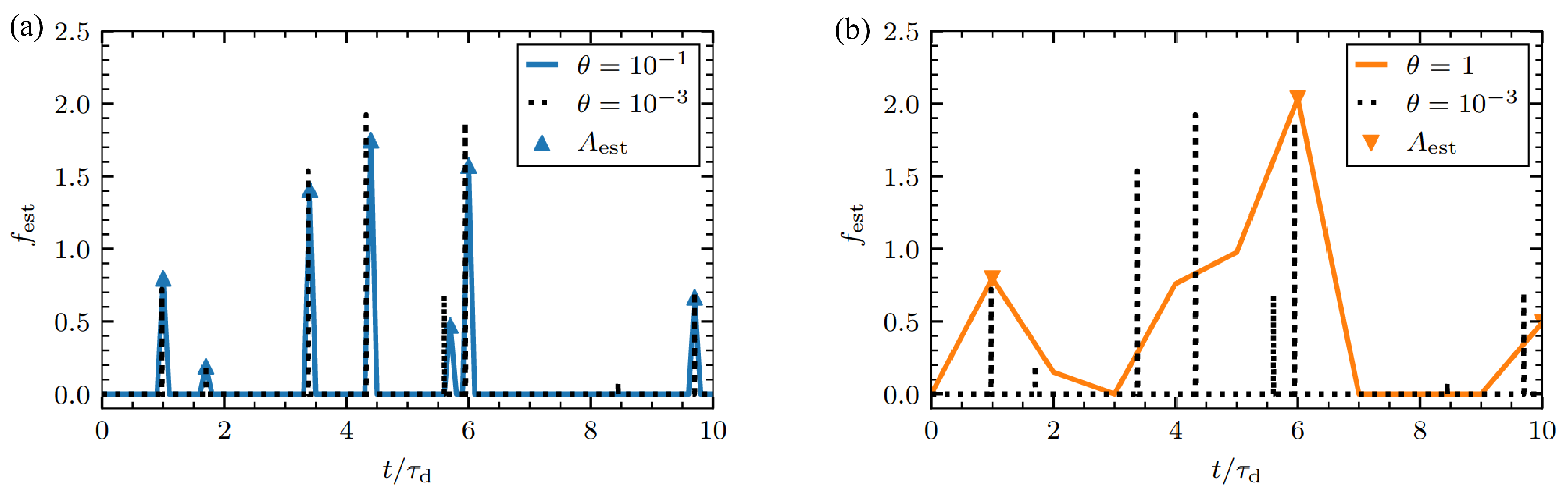}
    \caption{Estimated forcing $f_\mathrm{est}$ (solid line) and estimated amplitudes (markers) for $\gamma=1$ and two different normalized sampling times, (a) $\theta=10^{-1}$ and (b) $\theta = 1$. The original forcing (black dotted line) with $\theta=10^{-3}$ is shown for comparison.}
    \label{fig:f-est-sampling}
\end{figure}
\begin{figure}
    \centering
    \includegraphics[width=1.\textwidth]{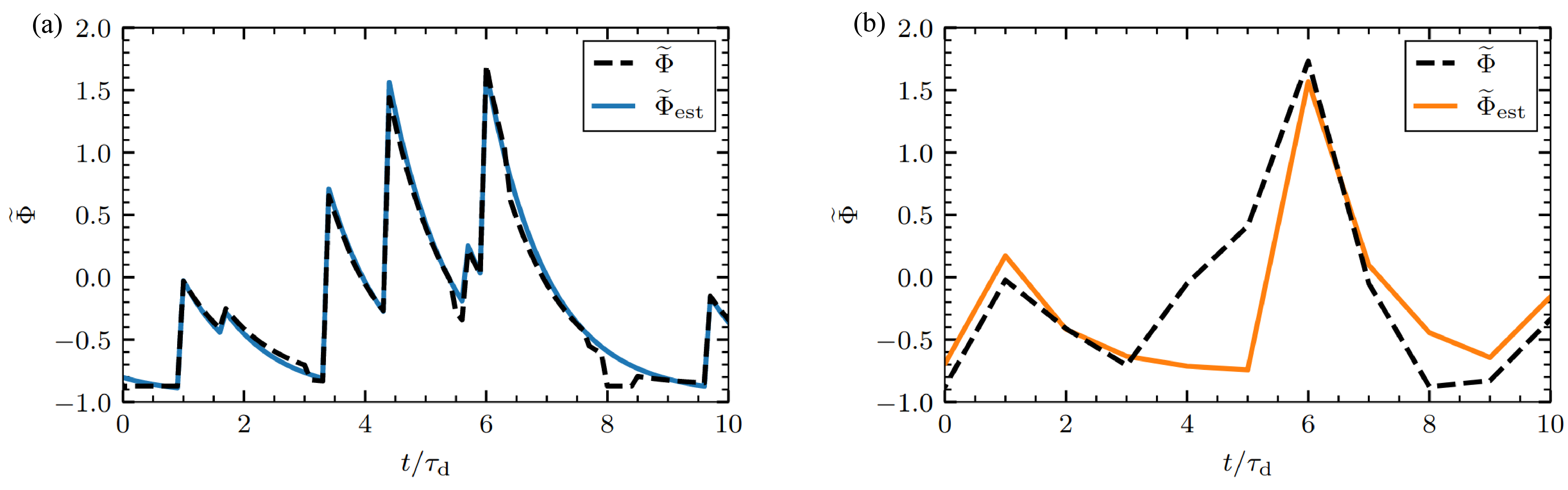}
    \caption{Comparison between reconstructed time series using the estimated amplitudes and arrival times (solid line) and downsampled original signals (dashed line) for $\gamma=1$ and two different normalized sampling times, (a) $\theta=10^{-1}$ and (b) $\theta = 1$.}
    \label{fig:recon-sig-est-amp-tw-sampling}
\end{figure}

The effects of varying the sampling time are presented in \Figsref{fig:f-est-sampling} and \ref{fig:recon-sig-est-amp-tw-sampling}. Here we keep $\gamma = 1$ fixed for various values of $\theta$. In these figures, a realization of the FPP with normalized sampling time $\theta = 10^{-3}$ was downsampled by using only a portion of the data points in the time series. The downsampled signals were then deconvolved with a similarly downsampled pulse function, and the signals were reconstructed. In \Figref{fig:f-est-sampling}, the estimated forcing is presented. As expected, increasing the sampling time leads to fewer arrivals identified as not all pulses can be separated. The result is similar to the effect of keeping $\theta$ fixed and increasing $\gamma$, as expected from the theoretical prediction presented in \Appref{app:number-events}.

In \Figref{fig:recon-sig-est-amp-tw-sampling} the reconstructed signals are compared to the downsampled original time series without noise, $X=0$. Note that the downsampled time series with $\theta = 10^{-1}$ is here visually similar to the original time series with $\theta = 10^{-3}$. The reconstruction is reasonable, even for $\theta = 1$, when compared to the downsampled signal. However, while the amplitudes and arrival times might be reasonably estimated in the case $\theta = 10^{-1}$, this is obviously not possible in the case $\theta = 1$, as expected.

\newpage
\section{Amplitude and waiting time distributions}\label{sec:rec-amp-wait}
In this section, we investigate the ability of the deconvolution algorithm to reconstruct the amplitude and waiting time distributions for the case where the pulse function is known. In all the following cases we use $\theta=10^{-2}$, $T/\td=10^{4}$ and we have not added any noise term. For this value of the sampling time, the condition $\gamma \theta \leq 1/20$ becomes $\gamma \leq 5$. The deconvolution ran for $10^{5}$ iterations, leading to convergence for $\gamma<10$ and marginal convergence for $\gamma \geq 10$.

\tikzset{ 
    table/.style={
        matrix of nodes,
        row sep=-\pgflinewidth,
        column sep=-\pgflinewidth,
        nodes={
            rectangle,
            draw=black,
            align=center
        },
        minimum height=1.5em,
        text depth=0.5ex,
        text height=2ex,
        nodes in empty cells,
        every even row/.style={
            nodes={fill=white}
        },
        column 1/.style={
            nodes={text width=7em}
        },
        row 1/.style={
            nodes={
                fill=white,
                text=black,
            }
        }
    }
}

\begin{figure}
\centering
\includegraphics[width=1.\textwidth]{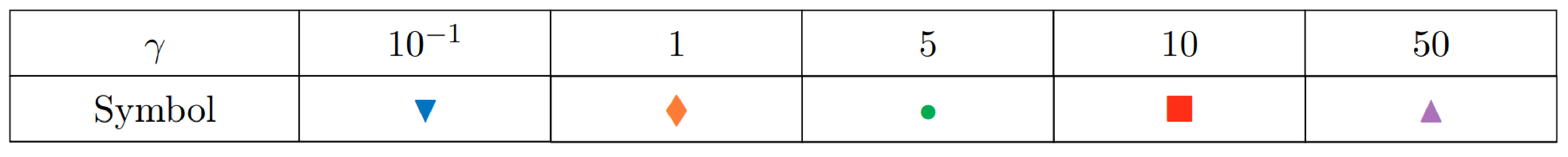}
\caption{List of plot symbols for various intermittency parameters used for estimating pulse amplitude and waiting time statistics.}
\label{tab:symbols-est-amp-tw}
\end{figure}


\subsection{Base case}

\begin{figure}
  \centering
  \includegraphics[width=1.\textwidth]{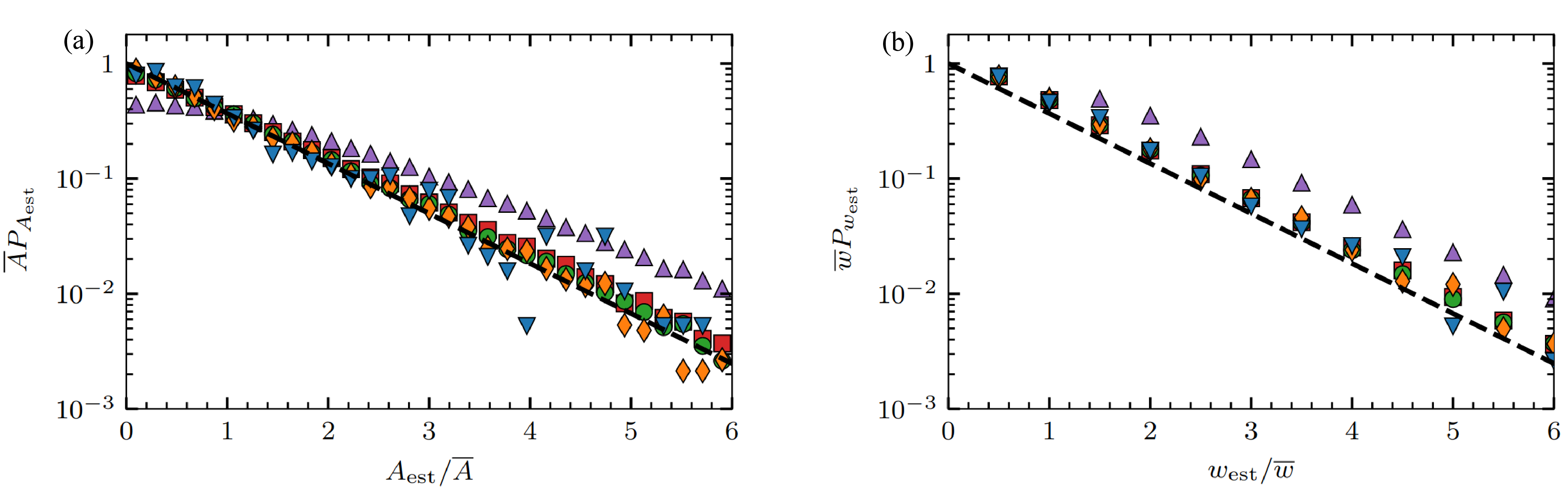}
  \caption{Probability distributions of (a) estimated amplitudes and (b) estimated waiting times for various intermittency parameters in the base case with exponentially distributed pulse amplitudes and waiting times. In both cases, the distributions are normalized by the original sample mean values. The plot symbols are defined in \Figref{tab:symbols-est-amp-tw}.}
  \label{fig:exp-amp-tw-dist}
\end{figure}

First, we consider the effects of varying $\gamma$ in the base case on the estimated amplitude and waiting time distributions from the deconvolved forcing. The plot symbols used for various intermittency parameters are presented in \Figref{tab:symbols-est-amp-tw}. The resulting distributions are presented in \Figref{fig:exp-amp-tw-dist} and the estimated parameters are presented in \Tabref{table:est-mean-exp}. Here, the found number of maxima, the estimated average amplitudes, and the estimated average waiting times are compared to the theoretical number of maxima in \Eqref{eq:approx-mean-M} and the theoretical mean values of the process. 
In the figures, the distributions are well estimated for $\gamma \leq 10$, while the mean values are within 10\% of their theoretic values for $\gamma \leq 5$. 
The expected number of maxima is well estimated by \Eqref{eq:approx-mean-M}.

For $\gamma=50$, the mean amplitude and waiting time are significantly overestimated compared to the other cases. For the amplitudes, this effect is mainly due to multiple events being added as a result of being closer than a sampling time. The resulting distribution has a shallower slope and is more concave when compared to the original distribution. For the waiting times, a significant number of true waiting times are below one sampling time (the probability $\text{Pr}[w<\theta]\approx0.39$ for $\gamma=50$). As the smallest waiting time resolvable by the deconvolution method is $2\theta$, this introduces a cutoff in the estimated distribution which increases the mean value of the estimated waiting times. Moreover, the waiting times only take values that are low-integer multiples of the sampling time. For consistency between different values of the intermittency parameter, we use the same bins for all distributions shown in these figures, chosen such that each bin contains a single integer multiple of the sampling time in the $\gamma=50$-case. 
Although it is elevated, due to discounting the shortest waiting times, the distribution for $\gamma=50$ decays with the same slope as the original distribution. 

In the following, we will investigate the robustness of the method to non-exponential amplitude and waiting time distributions. We have chosen to test the Rayleigh distribution due to its Gaussian tail, the Pareto distribution as an example of a much broader distribution than the exponential, the uniform distribution for its discontinuous cutoff towards large values, and the extreme case of the degenerate distribution. Definitions of these distributions are presented in \Appref{app:dist-def}.

\begin{table}
\begin{ruledtabular}
\centering
 \begin{tabular}{c||c c|c c c c c} 
  \multirow{2}{*}{} 
  & \multirow{2}{*}{Estimated averages} & & & & $\gamma$ & & \\
  & & & $10^{-1}$ & 1 & 5 & 10 & 50\\ [0.5ex] 
 \hline
  \multirow{3}{*}{\begin{tabular}{c}$A \sim \Exp$ \\ $w \sim \Exp$\end{tabular}} & $M$/$\langle M \rangle$ & &1.02 & 1.01 & 0.98 & 0.98 &0.95\\ 
   & $\Sm{A}_{\rm est}/\Sm{A}$ & &1.01 & 1.02 & 1.06 & \ldev{1.11} & \ldev{1.58}\\
   & $\Sm{w}_{\rm est}/\Sm{w}$ & & 1.01 & 1.03 & \ldev{1.10} & \ldev{1.19}& \ldev{2.03}\\
   [1ex]
 \end{tabular}\\
 \caption{Ratio of (top) number of maxima in the forcing estimated from the deconvolution to the theoretical number of maxima, (middle) mean estimated amplitudes to the original sample mean and (bottom) mean estimated waiting times to the original sample waiting times. Results are from the base case and for various intermittency parameters. 
 The corresponding distributions are presented in \Figref{fig:exp-amp-tw-dist}. 
 \label{table:est-mean-exp}}
\end{ruledtabular}
\end{table}



\subsection{Amplitude distributions}

\begin{figure}
  \centering
  \includegraphics[width=0.9\textwidth]{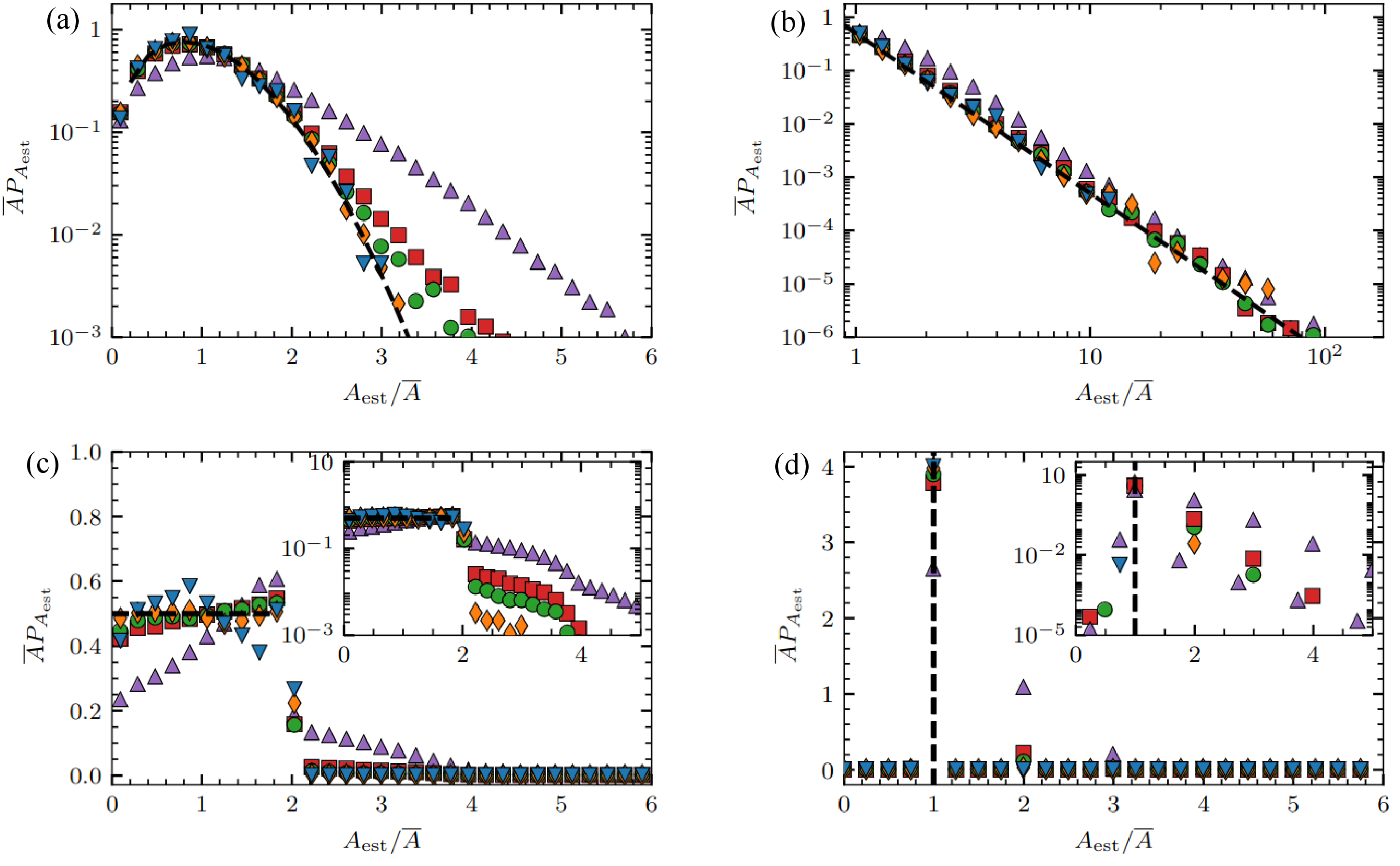}
  \caption{Probability distributions of estimated pulse amplitudes normalized by the sample mean amplitude for various intermittency parameters and distributions of pulse amplitudes in realizations of the process. (a) $A \sim \Ray$, (b) $A\sim \Par(3)$, (c) $A\sim \Unif$ and (d) $A\sim\Deg$. The insets in (c) and (d) show the distributions with semi-logarithmic scaling. In all cases, exponentially distributed waiting times were used. The black dashed lines represent the analytical amplitude distribution of the various realizations.
  The plot symbols are defined in \Figref{tab:symbols-est-amp-tw}.}\label{fig:amp-dist-var}
\end{figure}

In \Figref{fig:amp-dist-var} we present the amplitude distributions estimated from the deconvolution procedure for various pulse amplitude distributions and exponentially distributed waiting times. In all cases, the different symbols denote the estimated distributions corresponding to the intermittency parameter values given in \Figref{tab:symbols-est-amp-tw}. The black dashed line gives the analytical distributions. Note that the Rayleigh distribution is presented with a semi-logarithmic scale, the Pareto distribution is presented with a double-logarithmic scale and the uniform and degenerate distributions are presented with a linear scale, with a semi-logarithmic scale in the inset. \Tabref{table:amp-est-mean} 
aggregates mean values 
for the distributions in \Figref{fig:amp-dist-var}.

We see that all distributions are well estimated for $\gamma \leq 1$. The estimates for Rayleigh distributed amplitudes are visibly affected by pulse overlap for $\gamma \geq 5$, showing elevated and close to exponential tails for large amplitudes. Pareto distributed amplitudes are reliably identified for all tested intermittency parameters, likely due to the large range of probable values leading to little distortion due to pulse overlap. For uniformly distributed amplitudes, it is clear from the inset of \Figref{fig:amp-dist-var}(c) that while the implications of pulse overlap are visible even for $\gamma=1$, there is a jump of order the $10^{-2}$ from the originally allowed values to the larger values. As a consequence, the uniform distribution is well estimated and within the variation around the straight line expected for the uniform distribution. For a degenerate distribution of pulse amplitudes, the main contribution to the estimated amplitude distribution is the expected delta peak, with corrections at higher integer values, as seen in the inset. Only for $\gamma=50$ is there a significant contribution of normalized amplitudes larger than two and for non-integer amplitudes.

In \Tabref{table:amp-est-mean}, the number of found maxima is very close to the expected number corrected by effects of discretization and taking the 3-point maxima, as discussed in \Secref{sec:limitations-overlap}. This distortion is reflected in the deviation of the average estimated waiting time from the theoretical waiting time. There is only a small effect on the average amplitudes for $\gamma \leq 10$, irrespective of amplitude distribution. 

\begin{table}
\begin{ruledtabular}
\centering
 \begin{tabular}{c||c c|c c c c c} 
  \multirow{2}{*}{$P_A$} 
  & \multirow{2}{*}{Estimated averages} & & & & $\gamma$ & & \\
  & & & $10^{-1}$ & 1 & 5 & 10 & 50\\ [0.5ex] 
 \hline
  \multirow{3}{*}{$\Ray$} & $M$/$\langle M \rangle$  & & 1.03 & 1.02 & 1.00 & 1.00 &0.99\\ 
  & $\Sm{A}_{\rm est}/\Sm{A} $ & & 1.00 & 1.01 & 1.03 & 1.07 & \ldev{1.41}\\
  & $\Sm{w}_{\rm est}/\Sm{w}$ & & 1.00 & 1.02 & 1.08 & \ldev{1.16} & \ldev{1.95}\\[1ex]
 \hline
  \multirow{3}{*}{$\Par(3)$} & $M$/$\langle M \rangle$ & & 1.03 & 1.02 & 1.00 & 1.00 & 0.99\\ 
   & $\Sm{A}_{\rm est}/\Sm{A}$ & & 1.00 & 1.01 & 1.03 & 1.07 & \ldev{1.45}\\
    & $\Sm{w}_{\rm est}/\Sm{w}$ & & 1.00 & 1.02 & 1.08 & \ldev{1.16}& \ldev{1.95}\\[1ex]
\hline
  \multirow{3}{*}{$\Unif$} & $M$/$\langle M \rangle$ & & 1.02 & 1.01 & 0.99 & 0.99 & 0.97\\ 
   & $\Sm{A}_{\rm est}/\Sm{A}$ & & 1.00 & 1.01 & 1.04 & 1.08 & \ldev{1.46}  \\
    & $\Sm{w}_{\rm est}/\Sm{w}$ & & 1.01 & 1.03 &1.09 & \ldev{1.17} & \ldev{1.99}\\[1ex]
    \hline
  \multirow{3}{*}{$\Deg$} & $M$/$\langle M \rangle$ & & 1.03 & 1.02 & 1.00 & 1.00 & 0.96 \\ 
   & $\Sm{A}_{\rm est}/\Sm{A}$ & & 1.00 & 1.00 & 1.02 & 1.04 & \ldev{1.32}\\
    & $\Sm{w}_{\rm est}/\Sm{w}$ & & 1.00 & 1.02 & 1.08 & \ldev{1.16} & \ldev{2.01}\\[1ex]
 \end{tabular}
 \caption{In each row: Ratio of (top) found number of maxima in the forcing estimated from the deconvolution to the theoretical number of maxima, (middle) mean estimated amplitudes to the original sample mean and (bottom) mean estimated waiting times to the original sample waiting times. Results are for the various amplitude distributions used in \Figref{fig:amp-dist-var}, exponentially distributed waiting times and for various intermittency parameters.
 \label{table:amp-est-mean}}
\end{ruledtabular}
\end{table}

\subsection{Waiting time distributions}

\begin{figure}
  \centering
  \includegraphics[width=0.9\textwidth]{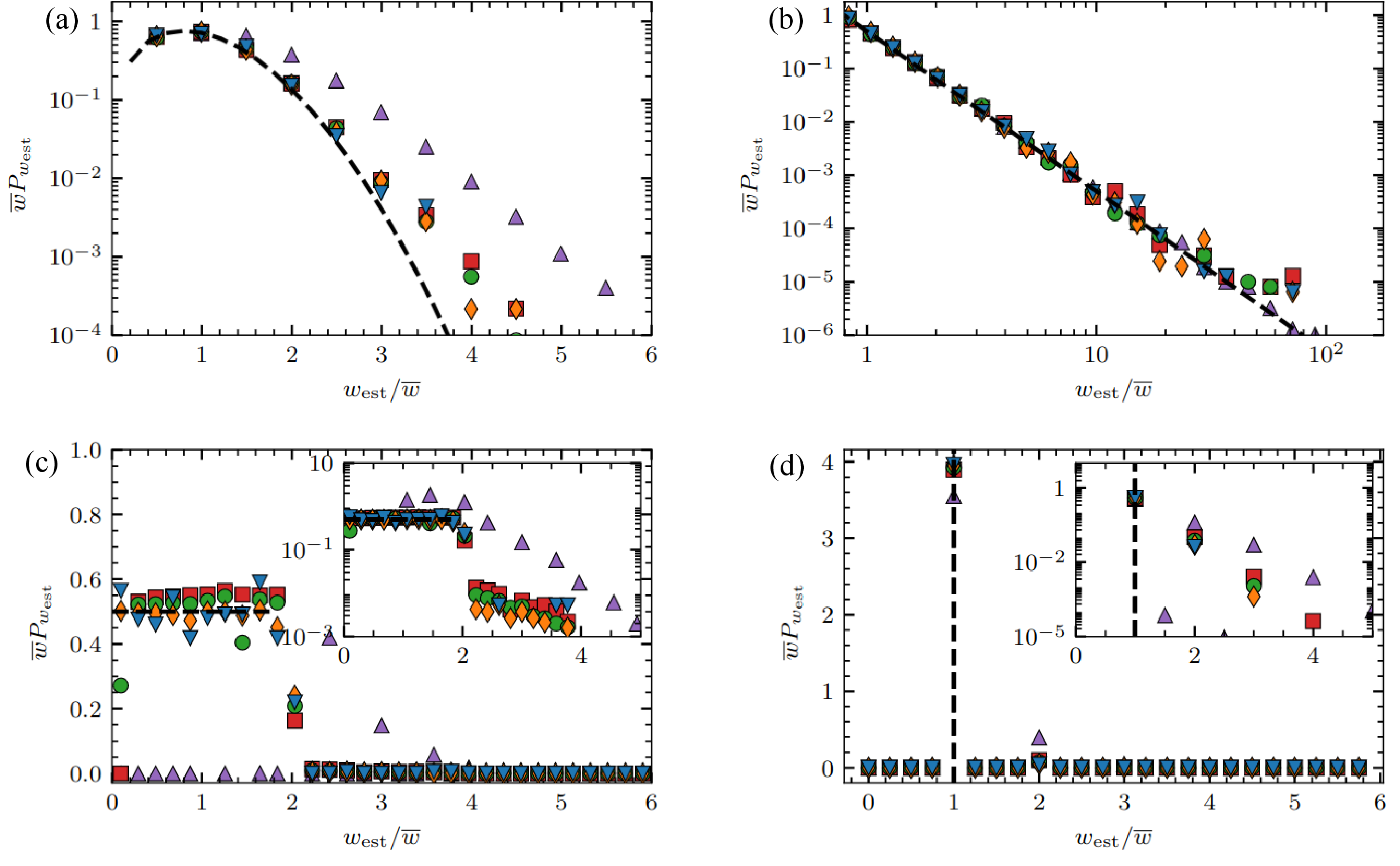}
  \caption{Probability distributions of estimated waiting times normalized by the sample mean waiting time for various intermittency parameters and distributions of waiting times in realizations of the process. (a): $w \sim \Ray$, (b): $w \sim \Par(3)$, (c): $w \sim \Unif$ and (d): $w \sim \Deg$. The insets in (c) and (d) show the distributions with semi-logarithmic scaling. In all cases, exponentially distributed amplitudes were used. The black dashed lines represent the analytical waiting time distribution of the various realizations.
  The plot symbols are defined in \Tabref{tab:symbols-est-amp-tw}.}\label{fig:tw-dist-var}
\end{figure}

Realizations of the process have also been made for various pulse waiting time distributions and intermittency parameters, with an exponential amplitude distribution in all cases. The estimated waiting time distributions from deconvolution of these realizations are presented in \Figref{fig:tw-dist-var} and mean amplitudes and waiting times as well as the number of maxima in the forcing are presented in \Tabref{table:tw-est-mean}. For $\gamma \leq 10$, all distributions show agreement with the true distribution for low ($w_\est < 3 \Sm{w}$) waiting times. The mean amplitudes and waiting times are well estimated for $\gamma \leq 10$, the largest deviation being a factor 1.11 increase for uniformly distributed waiting times and $\gamma=10$.

Rayleigh distributed waiting times display an elevated, exponential-like tail towards large values. Despite this, the mean values are better estimated than in the base case, possibly due to the clear peak of the Rayleigh distribution, ensuring most events are well separated. The distortion to the Rayleigh distribution for $\gamma=50$ follows the pattern seen for exponentially distributed waiting times, as discussed in the introduction to \Secref{sec:rec-amp-wait}. Again, we have chosen bins such that each bin contains one integer multiple of the sampling time in the case $\gamma=50$. We see that the tail of the distribution is elevated with respect to the original Rayleigh distribution and with respect to the lower $\gamma$ cases. The inflated mean values in \Tabref{table:tw-est-mean} are due to the loss of waiting times shorter than two time steps and corresponding enhanced pulse overlap.

Pareto distributed waiting times are identified in all cases, and the corresponding mean amplitudes and waiting times are almost perfectly estimated. This is due to pulse overlap not being a factor, even at these high intermittency levels: For $w\sim \Par(3)$, there is a cutoff $w/\td \geq 1/2\gamma$, see \Appref{app:dist-def}. With $\theta = 10^{-2}$, $\gamma=50$ corresponds to a minimal waiting time of 2 data points. There is an arrival at least every second data point, and this can just about be resolved by the 3-point maxima. A higher $\gamma$ would result in an artificial cutoff in the same manner as for the Rayleigh distributed waiting times.

For the degenerate distribution, $\theta=10^{-2}$ and $\gamma=50$ corresponds to a (degenerate) waiting time of two data points, again just at the edge of resolvability. Here, the (small) probability of larger waiting times is likely due to very small amplitudes not being picked up by the deconvolution or being removed by the $10^{-2}$ numerical noise threshold. As seen in \Tabref{table:tw-est-mean}, this has some effect on the average waiting time, but hardly any effect on the average amplitudes.

For the uniform distribution, the deconvolution cutoff $w/\Sm{w} \approx w/\Tm{w}  \geq 2 \Delta_t/\Tm{w} = 2 \theta \gamma$. In the $\gamma = 50$-case, this condition is $w/\Sm{w} \geq 1$ and is evident in \Figref{fig:tw-dist-var}(c): the probability of waiting times below this value is practically zero, and there is a corresponding positive probability for values larger than the original maximal value. The effect on the mean amplitude and waiting time is comparable to the effect in the case of Rayleigh distributed waiting times.

Comparing \Tabref{table:tw-est-mean} to \Tabref{table:est-mean-exp} we see that the theoretical $\Ttm{M}$ is only valid for exponentially distributed waiting times, although the uniformly distributed waiting times are within the $10\%$ margin for all $\gamma$. The most severe deviation is for Pareto distributed waiting times, where the number of events found is wrong by approximately a factor $1/\gamma$. This error is to be expected, as the exponential shape of the waiting time distribution, non-zero probability of evens closer than two time steps and the memory-less property of the Poisson process were central to the calculations in \Appref{app:number-events}. For the non-exponential waiting time distributions, at least two of these assumptions are false.

These results may indicate that the deconvolution procedure distorts amplitudes more than waiting times. Even if a cluster of arrivals is counted as a single arrival by the estimation algorithm, the time between such clusters is not distorted much by the algorithm, and the tail in the waiting time distribution should be well estimated, if there is no cutoff for large waiting times. The most significant source of error is the loss of events closer than two time steps.


\begin{table}
\begin{ruledtabular}
\centering
 \begin{tabular}{c||c c|c c c c c} 
  \multirow{2}{*}{$P_w$} 
  & \multirow{2}{*}{Estimated averages} & & & &  $\gamma$ & & \\
  & & & $10^{-1}$ & 1 & 5 & 10 & 50\\ [0.5ex] 
 \hline
  \multirow{3}{*}{$\Ray$} & $M$/$\langle M \rangle$ &  & 1.01 & 1.02 & 1.05 & \ldev{1.11} & \ldev{1.21}\\ 
  & $\Sm{A}_{\rm est}/\Sm{A}$ & & 1.01 & 1.01 & 1.01 & 1.02 & \ldev{1.28} \\
  & $\Sm{w}_{\rm est}/\Sm{w}$ & & 1.01 & 1.01 & 1.02 & 1.05 & \ldev{1.60}\\[1ex] 
 \hline
  \multirow{3}{*}{$\Par(3)$} & $M$/$\langle M \rangle$ &  & \ldev{9.76} & 1.01 & \ldev{0.21} & \ldev{0.11} & \ldev{0.04}\\ 
   & $\Sm{A}_{\rm est}/\Sm{A}$ & & 1.01 & 1.01 & 1.01 & 1.01 & 1.01\\
   & $\Sm{w}_{\rm est}/\Sm{w}$ & & 1.01 & 1.01 & 1.01 & 1.01 & 1.01\\[1ex]
   \hline
  \multirow{3}{*}{$\Unif$} & $M$/$\langle M \rangle$ & & 1.01 & 1.01 & 1.02 & 1.05 & 1.07\\ 
   & $\Sm{A}_{\rm est}/\Sm{A}$ & & 1.01 & 1.02 & 1.03 & 1.06 & \ldev{1.33}\\
    & $\Sm{w}_{\rm est}/\Sm{w}$ & & 1.01 & 1.02 & 1.06 & \ldev{1.11} & \ldev{1.64}\\[1ex]
    \hline
  \multirow{3}{*}{$\Deg$} & $M$/$\langle M \rangle$ & & 0.99 & 1.00 & 1.06 & \ldev{1.13} & \ldev{1.72}\\ 
   & $\Sm{A}_{\rm est}/\Sm{A}$ & & 1.01 & 1.01 & 1.01 & 1.01 & 1.03\\
    & $\Sm{w}_{\rm est}/\Sm{w}$ & & 1.01 & 1.01 & 1.02 & 1.03 & \ldev{1.13}\\[1ex]
 \end{tabular}
 \caption{Each row shows the ratio of (top) found number of maxima in the forcing estimated from the deconvolution to the theoretical number of maxima, (middle) mean estimated amplitudes to the original sample mean and (bottom) mean estimated waiting times to the original sample waiting times. Results are for the various waiting time distributions used in \Figref{fig:tw-dist-var}, exponentially distributed amplitudes and for various intermittency parameters.
 \label{table:tw-est-mean}}
 \end{ruledtabular}
\end{table}

\subsection{Negative pulse amplitudes}

\begin{figure}
  \centering
  \includegraphics[width=1.\textwidth]{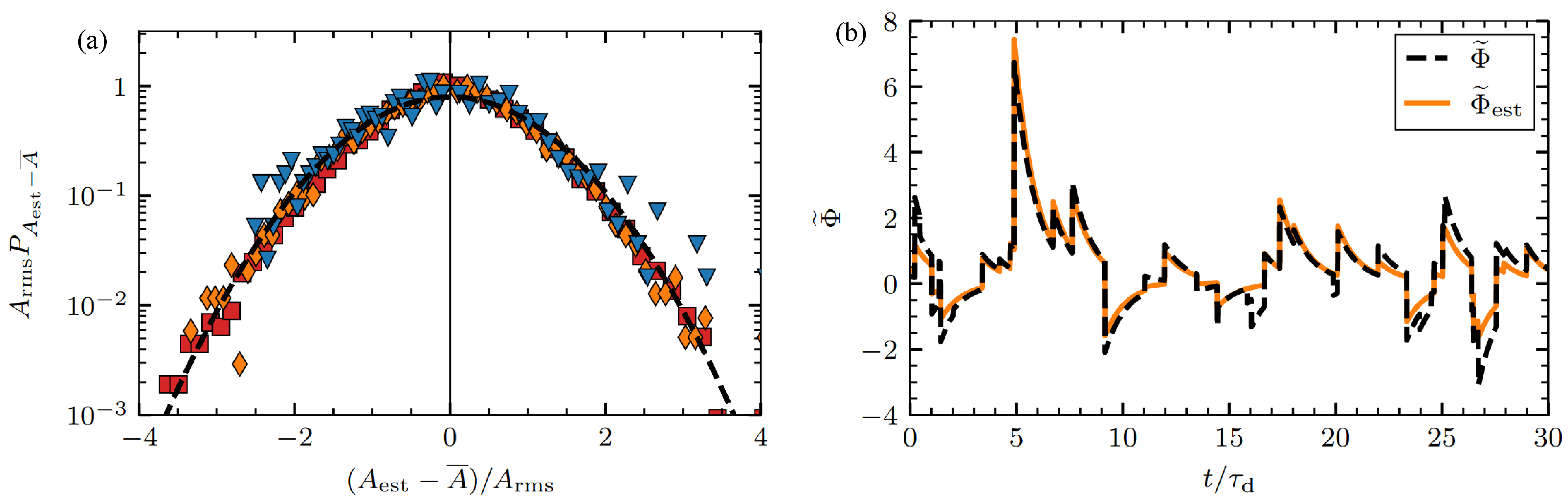}
  \caption{Reconstruction of a realization with normally distributed pulse amplitudes. (a) Probability distribution function of the amplitudes of found events from the deconvolution. Refer to \Figref{tab:symbols-est-amp-tw} for the legend. The analytical distribution (dashed line) is shown for reference, whereas the vertical solid lines show the separation between the negative found events and positive found events. (b) Reconstructed time series (solid line) using estimated amplitudes and arrival times for $\gamma=1$. The original time series (dashed line) is shown for comparison.}\label{fig:symmetric-amp-dist}
\end{figure}

All the amplitude distributions investigated above were positive definite. Now, we consider realizations with zero mean, normally distributed amplitudes to see if they are reproducible, and to investigate the robustness of the deconvolution to negative pulse amplitudes in the process. Symmetric Laplace distributed amplitudes were also tested with similar results but are not presented here. As the deconvolution algorithm works under the constraint that the forcing is non-negative, we first straightforwardly estimate the amplitudes and arrival times. This gives estimates of the positive amplitudes with corresponding arrival times. Then we multiply the signal by $-1$ and redo the deconvolution, giving estimates of the negative amplitudes with corresponding arrival times. In \Figref{fig:symmetric-amp-dist}, the results are presented for normally distributed amplitudes. In \Figref{fig:symmetric-amp-dist}(a), the resulting amplitude distributions are presented, where the vertical black line indicates the separation between positive and negative amplitudes. The overall shape of the normal distribution is well recovered. In \Figref{fig:symmetric-amp-dist}(b), an example of the reconstructed signal is compared to the original signal. It is clear that while large pulse amplitudes, both positive and negative, are identified, the method struggles for small amplitudes. Adding positive and negative pulses in quick succession leads to a signal shape which cannot be recognized by the deconvolution as a sum of just positive or just negative amplitudes, and this effect is more severe for small amplitudes.

\subsection{Conclusion}

In this most favorable case of a known pulse function and no noise added, the deconvolution method performs well overall, only limited by pulse overlap, as expected from the theory. Different distributions of pulse amplitudes and waiting times are well reproduced and mean amplitudes and waiting times are reliably estimated with an error of less than 10\% for intermittent signals, $\gamma \leq 5$, lending credence to the approximate condition for pulse recovery $\gamma \theta \leq 1/20$. Even signals with both positive and negative pulse amplitudes allow reconstruction of large amplitude events.

\newpage
\section{Effects of noise}\label{sec:noise}
In this section, we investigate the effects of adding normally distributed noise to the FPP in two different ways. \emph{Additive} noise consists of uncorrelated noise, while \emph{dynamical} noise has the same correlation function as the base case FPP, achieved by convolving uncorrelated noise with the pulse function. Some effects of these forms of noise, including methods for estimating parameters, are discussed further in Refs.~\onlinecite{theodorsen-ps-2017} and \onlinecite{theodorsen-ppcf-2018}. For both noise types, we use the base case FPP with uncorrelated, one-sided exponential pulses with exponentially distributed amplitudes. The noise to signal ratio is defined as $\eps = \rms{X}^2 / \rms{\Sn}^2$, where $\rms{X}$ is the standard deviation of the noise process. Here and in the following, $X_{\rm add}$ refers to additive noise, $X_{\rm dyn}$ refers to dynamic noise and $\Psi = \Sn + X$ refers to the process with noise.

\begin{figure}
    \centering
    \includegraphics[width=0.48\textwidth]{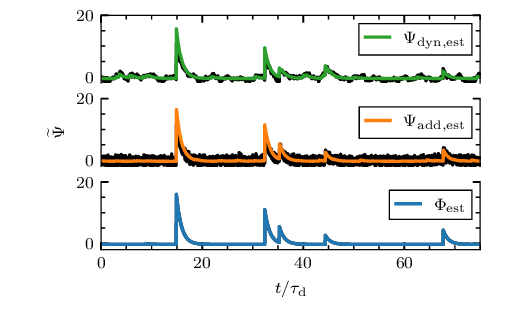}
    \caption{Original time series (black lines) and reconstructed time series shown by the colored (gray) lines from the deconvolution for no noise (lower panel), additive noise (middle panel) and dynamical noise (upper panel). The intermittency parameter was set to $\gamma=10^{-1}$, while the noise to signal ratio was set to $\epsilon = 1$.}
    \label{fig:noise-ts-comparison}
\end{figure}

\begin{figure}
    \centering
    \includegraphics[width=1.\textwidth]{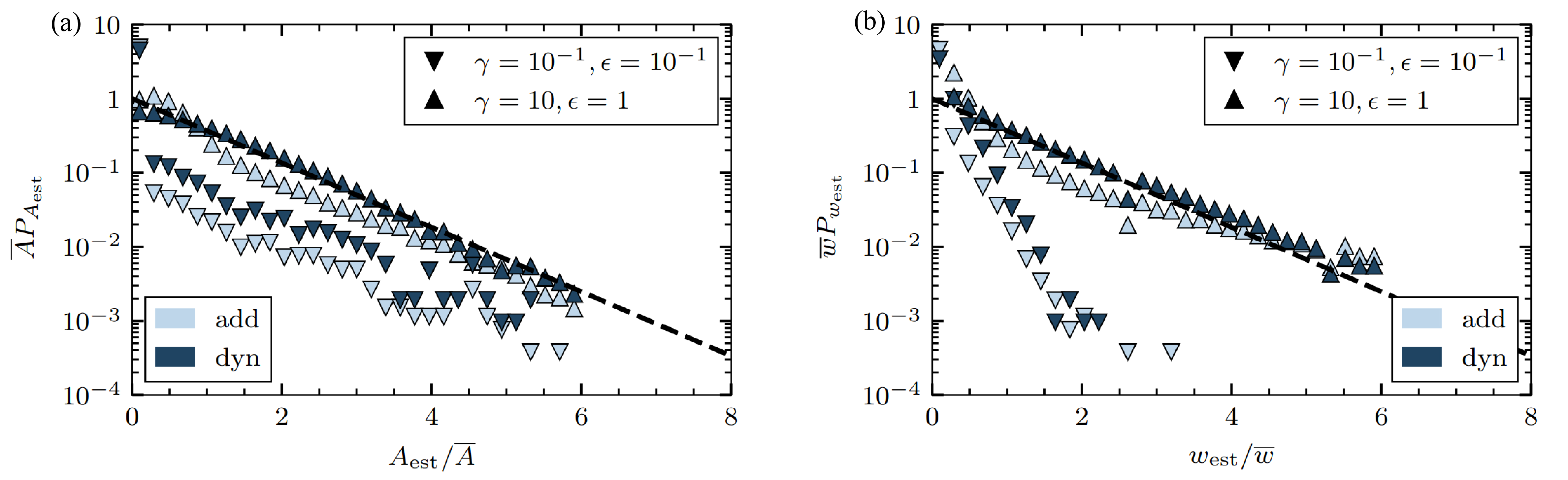}
    \caption{Probability distributions of (a) estimated amplitudes and (b) estimated waiting times without thresholding. The black dashed line gives an exponential distribution. The symbols represent the values of $\gamma$ and $\epsilon$ used, whereas the colors correspond to the noise type.}
    \label{fig:amp-tw-noise-no-thresh}
\end{figure}

In \Figref{fig:noise-ts-comparison} we present excerpts of the original and reconstructed time series for both types of noise and the case with no noise. In all cases, the same realization of the FPP was used. It is evident that additive noise, at least for the given noise variance and FPP intermittency, does not lead to major distortions in the reconstruction. This is expected, as the deconvolution algorithm takes normally distributed, additive noise into account and does not depend on the noise level. In \Figref{fig:noise-ts-comparison} a few small, spurious events can be seen, likely where the noise by chance approximately reproduces the pulse shape. For dynamical noise, we see more significant spurious events as the algorithm can no longer reliably separate the noise from the small amplitude pulses in the signal. This is seen in \Figref{fig:amp-tw-noise-no-thresh}, where estimated amplitude and waiting time distributions are presented, using only the small amplitude threshold discussed in \Secref{sec:extract-amp-arrivals} to avoid spurious arrivals. It is clear that many noise events are identified as pulses, and that this also influences the waiting time distribution. In all cases, the tail of the amplitude distribution follows the expected exponential decay. For low $\gamma$ and $\eps$, the effect of noise is largely concentrated in excessively many small-amplitude events which is reflected in the sharper decay of the waiting time distributions as compared to the original exponential distribution. For high $\gamma$ and $\eps$, the estimated amplitudes and waiting times are moderately affected, and in particular, dynamical noise has little effect on these distributions. In the following, we will discuss how to improve the results by removing the events with the smallest amplitudes.

\begin{figure}
    \centering
    \includegraphics[width=1.\textwidth]{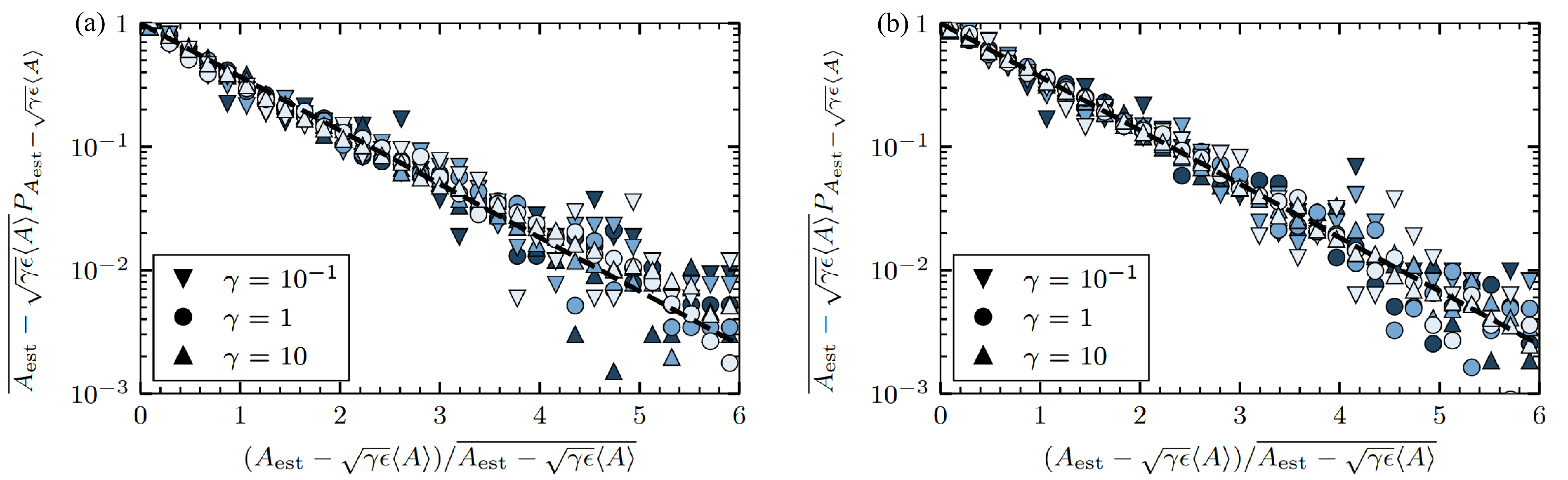}
    \caption{Probability distributions of estimated amplitudes with thresholding in the presence of (a) additive noise and (b) dynamical noise. The colored markers represents the different noise to signal ratios $\epsilon$, corresponding to different intermittency parameters $\gamma$ represented by the shape of the markers. Light blue (light gray) refers to $\eps = 10^{-1}$, medium blue (medium gray) refers to $\eps=1/2$ and dark blue (dark gray) represents $\eps=1$. The black dashed line is an exponential distribution.}
    \label{fig:amp-noise}
\end{figure}

\begin{figure}
    \centering
    \includegraphics[width=1.\textwidth]{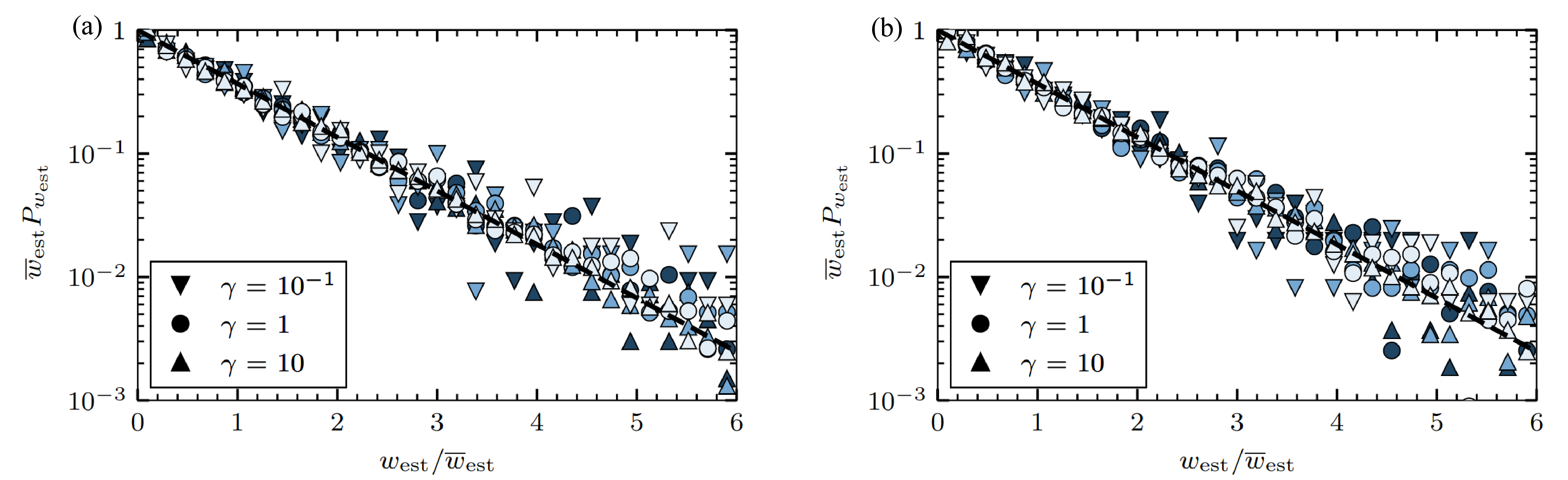}
    \caption{Probability distributions of estimated waiting times with thresholding in the presence of (a) additive noise and (b) dynamical noise. The colored markers represents the different noise to signal ratios $\epsilon$, corresponding to different intermittency parameters $\gamma$ represented by the shape of the markers. Light blue (light gray) refers to $\eps = 10^{-1}$, medium blue (medium gray) refers to $\eps=1/2$ and dark blue (dark gray) represents $\eps=1$. The black dashed line is an exponential distribution.}
    \label{fig:twdist-noise}
\end{figure}

Based on the exponential pulse amplitude and normal noise distributions, we have that $\rms{X} = \sqrt{\gamma \eps} \Ttm{A}$, so we introduce a threshold where we reject all events with amplitude less than this value. From realizations of the process, these parameters may be estimated from the moments, probability distribution or characteristic function as described in Refs.~\onlinecite{theodorsen-ps-2017} and~\onlinecite{theodorsen-ppcf-2018}. To simplify the analysis, we take $\gamma$, $\epsilon$ and $\Ttm{A}$ as given for setting the threshold level. In \Figsref{fig:amp-noise} and \ref{fig:twdist-noise} we present estimated amplitude and waiting time distributions after applying the amplitude criterion $A_\est > \sqrt{\gamma \eps}\Ttm{A}$, discarding all events with smaller amplitudes. Note that for these figures, we normalize by the sample mean of the estimated values instead of the sample mean of the original values to highlight the similarity in distribution. The result is that we have good agreement in distribution in all cases of pulse overlap and noise levels investigated. Removing the arrivals due to noise also realigns the waiting time distributions to the expected exponential.

In Tabs.~\ref{table:amp-est-var-eps-gamma} and~\ref{table:tw-est-var-eps-gamma-rescaled} the estimated mean amplitudes and waiting times for all cases presented in \Figsref{fig:amp-noise} and \ref{fig:twdist-noise} are presented. To accurately assess mean values of amplitudes and waiting times, the threshold must be taken into account. The mean value of the truncated amplitudes is, assuming exponentially distributed amplitudes, just the threshold subtracted from the sample mean value of the estimated amplitudes. Therefore, we report $\left(\Sm{A}_{\rm est}-\sqrt{\gamma\epsilon}\Tm{A}\right)/\Sm{A}$.  For waiting times, pulses are rejected if they have an amplitude $A<\sqrt{\gamma \epsilon} \Ttm{A}$, thus the number of found pulses after thresholding is reduced by a factor $1-\text{Pr}[A<\sqrt{\gamma \epsilon} \Ttm{A}] = \exp(-\sqrt{\gamma \eps})$. Here, the last equality holds for exponentially distributed amplitudes. This in turn implies that the estimated waiting time is prolonged and should be multiplied by this same factor, leading us to report $\exp(-\sqrt{\gamma \eps}) \Sm{w}_{\rm est}/\Sm{w}$. Note that this argument relies on an assumption of no pulse overlap and will be less accurate for intermittency parameters of order unity or larger.

In the case of the amplitudes, presented in \Tabref{table:amp-est-var-eps-gamma}, we recover the mean value well when we correct for the threshold used, although additive noise leads to underestimation of the amplitudes while dynamic noise leads to overestimation of amplitudes in most cases investigated. The estimated waiting times, presented in \Tabref{table:tw-est-var-eps-gamma-rescaled}, are in general much more affected by the noise than the amplitude distribution, reaching close to twice the original sample mean value. In both cases, it appears like moderate pulse overlap, $\gamma=1$, is more affected by noise than either low or high pulse overlap. It may be that for low pulse overlap, noise is effectively filtered out, while for high degrees of pulse overlap, the threshold is so restrictive (see below) that only the largest amplitude pulses are identified. We also recall that $\Sm{w}_{\rm{est}}/\Sm{w} = 1.19$ is expected for $\gamma=10$. Thus, the errors due to noise dominate for low and moderate $\gamma$, while for high $\gamma$, the errors due to pulse overlap are more significant.

\begin{figure}
    \centering
    \includegraphics[width=1.\textwidth]{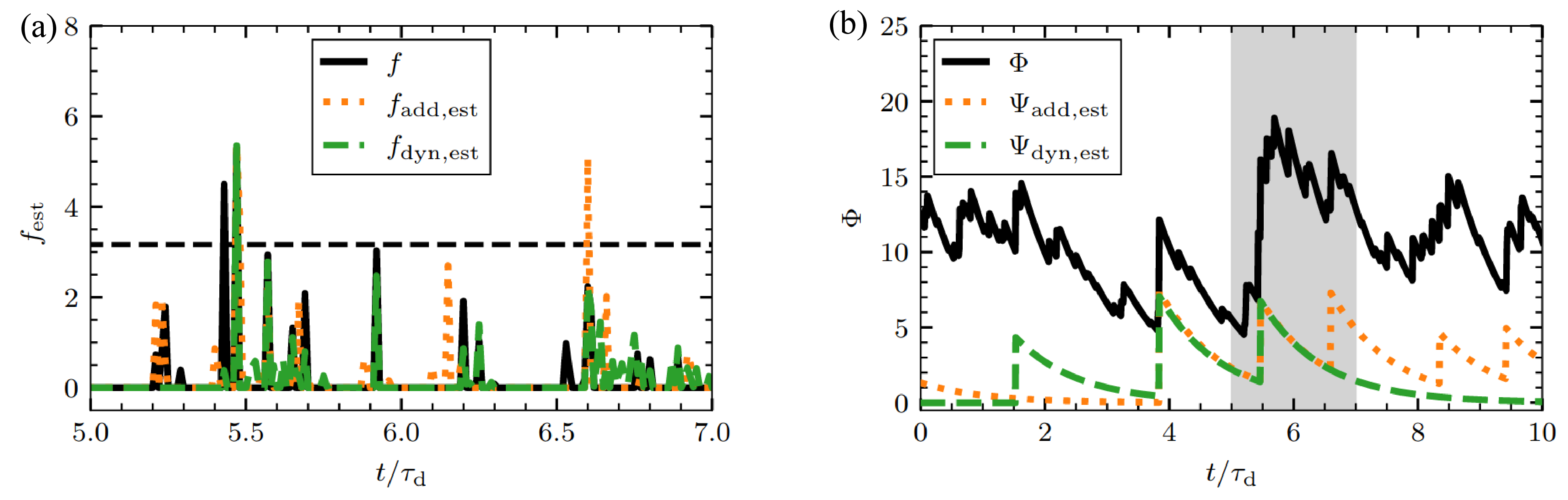}
    \caption{(a) Excerpt of the original forcing (black solid line) with $\gamma=10$ and $\eps=1$, compared to the estimated forcing using additive noise (orange dotted line) and dynamical noise (green dashed line) where the noise threshold is shown (horizontal black dashed line). The reconstructed time series (b) is shown using the estimated amplitudes and arrival times from the 3-point maxima. The shaded background corresponds to the time axis shown on the left.}\label{fig:reconstruct-sig-noise}
\end{figure}

In \Figref{fig:reconstruct-sig-noise}, we present the reconstructed time series from the deconvolution in the most extreme case considered, $\gamma=10$ and $\eps=1$. The original forcing (without noise) is compared to the forcing estimated from the signal with additive noise (orange dotted lines) and dynamical noise (green dashed lines) in \Figref{fig:reconstruct-sig-noise}(a). The horizontal black dashed line gives the amplitude threshold $\sqrt{\gamma \eps}$. It is clear that many small amplitude pulses are rejected as a result of the thresholding. Also note that additive noise appears to more severely affect large amplitudes than dynamical noise. In particular, note the large, spurious peak at normalized time 6.6. The reconstructed signals using the estimated forcing are compared to the original signal without noise in \Figref{fig:reconstruct-sig-noise}(b). It is clear that since only the largest amplitude pulses are identified, the method fails to reconstruct realizations of the process.

We have shown that the deconvolution recovers the forcing well in the case of normally distributed noise. To estimate amplitudes and arrival times, an amplitude threshold must be introduced. For high degrees of pulse overlap, this threshold significantly affects signal reconstruction, but by properly taking the threshold into account, we may recover the tails of the amplitude and waiting time distributions as well as estimating their mean values. The mean amplitude is accurately recovered, while the mean of the estimated waiting times is within a factor two of the original sample waiting time, even for severe noise levels.

\begin{table}
\begin{ruledtabular}
\centering
 \begin{tabular}{c||c c|c c c} 
 & \multirow{2}{*}{$\epsilon$} & &\multicolumn{3}{c}{$\gamma$} \\
   &  & & $10^{-1}$  & 1 & 10 \\ [0.5ex] 
 \hline
  \multirow{3}{*}{Additive} & $10^{-1}$ & & 0.90 & 0.90 & \ldev{0.89}\\ 
   & 1/2 & & 0.90 & \ldev{0.87} & \ldev{0.89}\\
   & 1 & & 0.92 & \ldev{0.84} & 0.94  \\[1ex]
 \hline
  \multirow{3}{*}{Dynamic} & $10^{-1}$  & & 1.01 & 1.03 & 1.07 \\
  & 1/2 & & 1.00 & 1.02 & 1.05\\
  & 1 & & 0.98 & 1.02 & 1.04 \\[1ex]
 \end{tabular}
 \caption{Estimated average amplitudes in the case of amplitude noise thresholding corrected to $\left(\Sm{A}_{\rm est}-\sqrt{\gamma\epsilon}\Tm{A}\right)/\Sm{A}$ corresponding to \Figref{fig:amp-noise}.\label{table:amp-est-var-eps-gamma}}
 \end{ruledtabular}
\end{table}


\begin{table}
\begin{ruledtabular}
\centering
 \begin{tabular}{c||c c|c c c} 
 & \multirow{2}{*}{$\epsilon$} & & \multicolumn{3}{c}{$\gamma$} \\
   &  & & $10^{-1}$  & 1 & 10 \\ [0.5ex] 
 \hline
  \multirow{3}{*}{Additive} & $10^{-1}$ & & 1.02 & \ldev{1.24} & \ldev{1.36}\\ 
   & 1/2 & & \ldev{1.17} & \ldev{1.63} & \ldev{1.36}\\
   & 1 & & \ldev{1.32} & \ldev{1.89} & \ldev{1.22}  \\[1ex]
 \hline
  \multirow{3}{*}{Dynamic} & $10^{-1}$ & & \ldev{1.12} & \ldev{1.25} & \ldev{1.25}\\ 
   & 1/2 & & \ldev{1.25} & \ldev{1.55} & \ldev{1.41}\\
   & 1 & & \ldev{1.37} & \ldev{1.82} & \ldev{1.52}  \\[1ex]

 \end{tabular}
 \caption{Estimated average waiting times in the case of amplitude noise thresholding corrected to $\exp(-\sqrt{\gamma \eps}) \Sm{w}_{\rm est}/\Sm{w}$ corresponding to \Figref{fig:twdist-noise}.\label{table:tw-est-var-eps-gamma-rescaled}}
 \end{ruledtabular}
\end{table}

\newpage
\section{Estimating the pulse duration}\label{sec:td-est}
For any given measurement time series, the pulse function and duration may not be known and so must be estimated in order to apply the deconvolution method. For pulses with fixed duration arriving in accordance to a Poisson process, the pulse function can be obtained from the frequency PSD \cite{garcia-pop-2017-2}. Alternatively, the conditionally averaging method for large-amplitude events may be used in the case of weak pulse overlap \cite{garcia-nme-2017,theodorsen-pop-2018,garcia-pop-2018-3}.

In this section, we investigate how different pulse duration and waiting time distributions distort the estimated pulse function and thereby the estimated amplitude and waiting time distributions. We will estimate the average duration from the frequency power spectra of the process, which, in contrast to the auto-correlation function, are very robust to a distribution of pulse durations \cite{garcia-pop-2017-2}. We restrict ourselves to the one-sided exponential pulse function, as it has a clearly identifiable Lorentzian power spectrum. We further restrict ourselves to deviations from the base case which produce reasonable Lorentzian-like power spectra. This means that degenerate and uniform waiting time distributions are not considered, as they produce pronounced peaks in the spectra. Pareto distributed waiting times produce mild deviations in the tail, which are considered acceptable. Exponentially distributed pulse durations will not be considered due to the significantly increased zero-frequency value in the PSD \cite{garcia-pop-2017-2}, and neither will Pareto distributed durations due to their drastic effects on the power spectra \cite{Korzeniowska_unpub}. Both uniformly distributed and Rayleigh distributed durations give Lorenzian-like power spectra. To include a narrow (compared to the Rayleigh), unimodal distribution that is still positive definite, we use a Gamma distribution with shape parameter $20$.

\begin{figure}
    \centering
    \includegraphics[width=1.\textwidth]{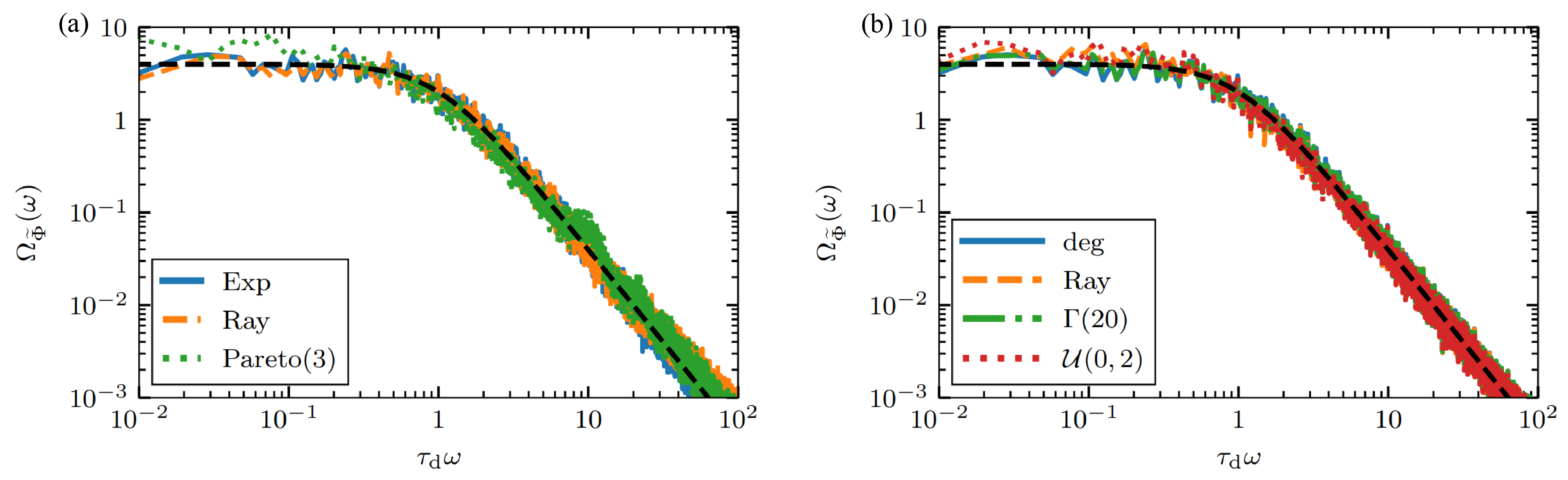}
    \caption{Power spectral densities of normalized original time series for $\gamma=10$ with (a) fixed pulse duration and different waiting time distributions and (b) exponential waiting time distribution and different pulse duration  distributions. The black dashed line is the Lorentzian spectrum for the base case.}
    \label{fig:psd-tw-td-dist}
\end{figure}

In \Figref{fig:psd-tw-td-dist}, we present the PSD of the normalized synthetic time series without added noise, for the various selected distributions of durations and waiting times. The intermittency parameter has very little visible effect on the power spectra, so figures for other intermittency parameters are not presented. Indeed, for the base case with a degenerate distribution of pulse durations and an exponential waiting time distribution, the PSD of the normalized process $\widetilde{\Phi}$ does not depend on the intermittency parameter \cite{garcia-pop-2017-2}. The theoretical expectation for the base case is apparently very close to the spectra for all cases presented in \Figref{fig:psd-tw-td-dist}. This will be quantified by estimates of the average duration in \Secsref{sec:td-est-wait-dist} and \ref{sec:td-est-td-dist}.

\subsection{Wrongly estimated pulse duration}

Before investigating the effect of a distribution of pulse durations and waiting times on the deconvolution method, we consider the isolated effect of misidentifying the correct duration for the base case process. Here, we deconvolve a base case realization with a pulse function with a wrong duration time in the case of no noise added. We will keep the notation $\td=\Ttm{\tau}$ for the true duration of the process and use $\tda$ for the assumed wrong duration. 

\begin{figure}
    \centering
    \includegraphics[width=1.\textwidth]{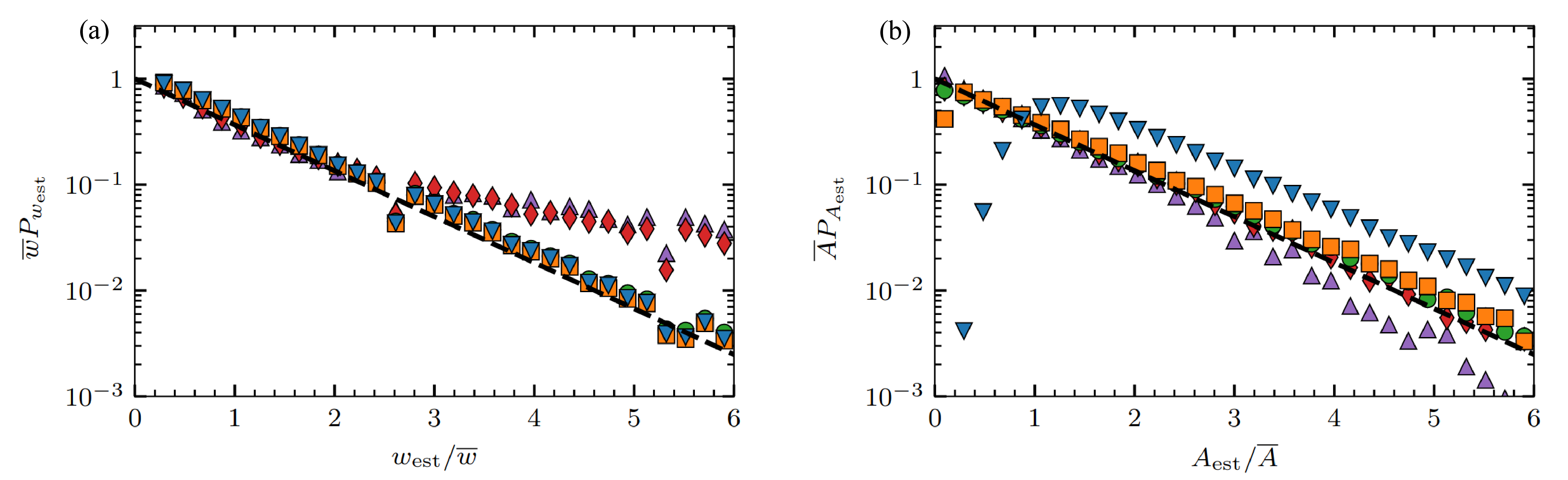}
    \caption{Probability distribution function of estimated (a) waiting times and (b) amplitudes for $\gamma=10$ and various assumed values of $\tda$.  The black dashed lines shows an exponential distribution. $\color{RoyalBlue}{\blacktriangledown}$ represents $\tda/\tau_\mathrm{d} = 1/10$, $\color{Orange}{\blacksquare}$ shows the data for $\tda/\tau_\mathrm{d} = 1/2$, $\color{Green}{\bullet}$ shows the data for $\tda/\tau_\mathrm{d} = 1$ $\color{Red}{\blacklozenge}$ shows the data for $\tda/\tau_\mathrm{d} = 2$ and $\color{Orchid}{\blacktriangle}$ represents $\tda/\tau_\mathrm{d} = 10$.}
    \label{fig:td-wrong-recon}
\end{figure}

In \Figref{fig:td-wrong-recon}, the estimated amplitude and waiting time distributions are presented for various assumed $\tda$. The intermittency parameter for all realizations is $\gamma=10$. Similar results are present, but with a weaker effect, for $\gamma=1$. We observe deviations from the exponential distribution of waiting times when $\tda > \td$ and deviations for an exponential amplitude distribution for the two most extreme cases with $\tda/\td=1/10$ and $\tda/\td=10$. This is complemented by tables \ref{tab:mean_est_tw_diff_td} and \ref{tab:mean-est-amp-diff-td} where the estimated average waiting time and amplitude are compared to their respective original sample mean values. Large $\tda/\td$ leads to overestimation of the average waiting time and this is more pronounced for larger intermittency parameters. Large $\tda/\td$ also leads to underestimation of the average amplitude and this is most pronounced for small intermittency parameters. Small $\tda/\td$ has very little effect on the average waiting time but leads to overestimation of the average amplitude, in particular for large intermittency parameters.

We interpret these results as follows. The deconvolution preserves the integral of the signal (at least before the 3-point maxima is applied) and the integral of the pulse function is equal to the duration time. Therefore, overestimating or underestimating the pulse duration leads to decreased or increased mass in the estimated forcing, respectively. Increased mass in the forcing raises the zero-level of the entire forcing, increasing the amplitudes but not causing any events to be lost. Decreased mass in the forcing can only be achieved by decreasing pulse amplitudes, which also eliminates some pulses entirely. Hence, overestimating the duration time leads to lost pulse arrivals, while underestimating the duration mainly leads to increased pulse amplitudes, which is moderate unless the underestimation is extreme or there is significant pulse overlap.  These effects suggest that in application of the deconvolution method, where the pulse duration must be estimated, one should favor doing the deconvolution with a slightly lower duration than the estimated one. Noise, different distributions of amplitudes, waiting times, or durations, or correlations between these random variables could change this conclusion, however.

\begin{table}
\begin{ruledtabular}
\centering
 \begin{tabular}{c||c c c c c c c c} 
\multirow{2}{*}{$\gamma$} & \multicolumn{8}{c}{$\tda/\td$} \\
   &   $10^{-1}$ & 1/2 & 1 & 11/10 & 5/4 & 3/2 & 2 & 10\\ [0.5ex] 
 \hline
   $10^{-1}$ & 1.01 & 1.01 & 1.01 & 1.02 & 1.04 & 1.06 & \ldev{1.12} & \ldev{1.78}\\ 
  1 & 1.02 & 1.01 & 1.03 & 1.09 & \ldev{1.18} & \ldev{1.33} & \ldev{1.60} & \ldev{4.53}\\ 
  10 & \ldev{1.17} & \ldev{1.16} & \ldev{1.19} & \ldev{1.28} & \ldev{1.44} & \ldev{1.71} & \ldev{2.22} & \ldev{9.22}\\
 \end{tabular}
\caption{Table of the estimated average pulse waiting times normalized by the mean of the sample waiting times, $\Sm{w}_{\est}/\Sm{w}$, for different pulse durations at different intermittency values.}
\label{tab:mean_est_tw_diff_td}
\end{ruledtabular}
\end{table}

\begin{table}
\begin{ruledtabular}
\centering 
 \begin{tabular}{c||c c c c c c c c} 
\multirow{2}{*}{$\gamma$} & \multicolumn{8}{c}{$\tda/\td$} \\
   &   $10^{-1}$ & 1/2 & 1  & 11/10 & 5/4 & 3/2 & 2 & 10\\ [0.5ex] 
 \hline
  $10^{-1}$ & 1.02 & 1.01 & 1.01 & 0.97 & 0.91 & \ldev{0.82} & \ldev{0.70} & \ldev{0.25}\\ 
  1 & 1.10 & 1.01 & 1.02 & 1.00 & 0.97 & 0.93 & \ldev{0.86} & \ldev{0.50}\\ 
  10 & \ldev{1.97} & \ldev{1.20} & \ldev{1.11} & \ldev{1.11} & 1.10 & 1.09 & 1.07 & \ldev{0.88}\\
 \end{tabular}
\caption{Table of the estimated average pulse amplitudes normalized by the sample mean amplitudes, $\Sm{A}_\est/ \Sm{A}$, for different pulse durations at different intermittency values.}
\label{tab:mean-est-amp-diff-td}
\end{ruledtabular}
\end{table}

\subsection{Effect of waiting time distribution}\label{sec:td-est-wait-dist}

Here, we will consider how various waiting time distributions affect the estimated pulse duration, and in turn how this influences the estimated amplitude and waiting time statistics using the deconvolution method. In \Tabref{tab:params-tw-dist-est}, the estimated pulse duration for the cases in \Figref{fig:psd-tw-td-dist}(a) are presented. These estimated durations were found by performing a least-square minimization using the curve\_fit function of the SciPy module in Python, in the normalized frequency range $\td\omega$ between $10^{-1}$ to $10^2$. The pulse duration is well estimated for the case of exponentially and Rayleigh distributed waiting times. The pulse amplitude and waiting time distributions estimated from the deconvolution method are not affected by the duration estimate in these cases, and are therefore not presented.

\begin{table}
\begin{ruledtabular}
\centering
 \begin{tabular}{c||c c c} 
\multirow{2}{*}{$\tde/\td$} &  \multicolumn{3}{c}{$\gamma$} \\
      &  $10^{-1}$ & 1  & 10 \\ [0.5ex] 
 \hline
  $w \sim \Exp$ & 1.00 & 1.08 & 0.96\\ 
  $w \sim \Ray$ & 0.93 & 1.04 & 0.99 \\ 
  $w \sim \Par(3)$ & \ldev{1.51} & \ldev{1.41} & \ldev{1.42} \\ 
 \end{tabular}
\caption{Ratio between estimated pulse duration $\tde$ and the pulse $\td$ for different waiting time distributions, corresponding to the spectra presented in \Figref{fig:psd-tw-td-dist}(a).}
\label{tab:params-tw-dist-est}
\end{ruledtabular}
\end{table}

\begin{figure}
    \centering
    \includegraphics[width=1.\textwidth]{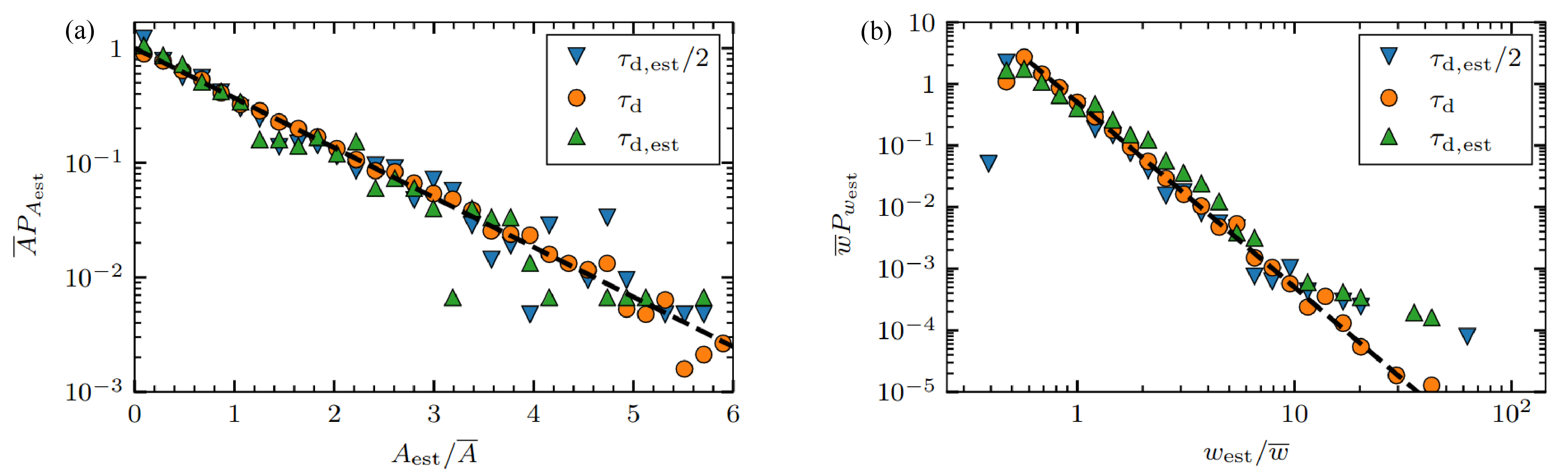}
    \caption{Estimated (a) amplitude distributions and (b) waiting time distributions for Pareto distributed waiting times, $w\sim\Par(3)$, and $\gamma=10^{-1}$. These plots compare the results from deconvolution performed using the true duration (orange circles) and the estimated duration (green triangles) from  \Tabref{tab:params-tw-dist-est}. The black dashed lines are the references of the input distributions.}
    \label{fig:td-dist-tw-dist}
\end{figure}

Pareto distributed waiting times gives an overestimation of the pulse duration, and we expect this to have effects on the subsequently estimated amplitude and waiting time distributions using the deconvolution method. The case $\gamma=10^{-1}$ has the largest overestimation and will therefore be investigated in more detail. In \Figref{fig:td-dist-tw-dist}, we present estimated amplitude and waiting time distributions from deconvolution for pulse durations $\td$, $\tde$ and $\tde/2$. The first is included as a baseline reference, while the third is included to demonstrate the effect of using a shorter duration than the estimated one. The amplitude distribution is not visibly affected, but the estimated waiting time distributions show elevated tails at large waiting times ($>10\Sm{w}$) and in the case where we used $\tde$, the distribution is elevated for waiting times greater than $\Sm{w}$.

The mean values of all distributions presented in \Figref{fig:td-dist-tw-dist} are presented in \Tabref{tab:params-est-twdist-pareto}. The mean amplitudes are consistently underestimated, both for $\tde>\td$ and for $\tde/2 < \td$, in contrast to the results in \Tabref{tab:mean-est-amp-diff-td}. We see that the deconvolution significantly overestimates $\Sm{w}_\est$ for $\tde$, and underestimates $\Sm{w}_\est$ for $\tde/2$. In agreement with the results reported in the previous section, using a slightly lower pulse duration than the estimated one improves the estimated average amplitude and waiting time.

\begin{table}
\begin{ruledtabular}
\centering
 \begin{tabular}{c||c c c}
   Estimated averages &   $\td$ & $\tde/2$  & $\tde$ \\ [0.5ex] 
 \hline
  $\Sm{A}_{\rm est}/\Sm{A}$ & 1.01 & 0.91 & \ldev{0.88}\\ 
  $\Sm{w}_{\rm est}/\Sm{w}$ & 1.01 & 0.91 & \ldev{1.28}\\ 
 \end{tabular}
\caption{Comparison of rescaled estimated parameters from \Figref{fig:td-dist-tw-dist} for Pareto waiting times, $\gamma = 10^{-1}$, using different duration time values in the deconvolution.}
\label{tab:params-est-twdist-pareto}
\end{ruledtabular}
\end{table}

\subsection{Distribution of pulse durations}\label{sec:td-est-td-dist}

Consider now the situation where there is a distribution of pulse durations. The deconvolution method assumes all pulses to have the same duration, so it is of interest to investigate how it performs for a distribution of pulse durations. We first establish a baseline for the performance of the method using the theoretical average duration time. In \Tabref{table:params-td-dist-amp-tw}, we present estimated average amplitudes and waiting times for three different pulse duration distributions. Table~\ref{table:est-mean-exp} describes the results for a degenerate distribution of pulse durations. While Gamma distributed durations with shape parameter $20$ is narrow enough for the parameters to be well estimated, in both the Rayleigh and uniform cases, the amplitudes are underestimated for all intermittency parameters. For small intermittency parameters, this agrees with the results in \Tabref{tab:mean-est-amp-diff-td}: the amplitudes of pulses with duration greater than $\td$ are accurately reconstructed, while amplitudes of pulses with durations smaller than $\td$ are underestimated. Pulse overlap modifies this relationship, leading to more robustly underestimated amplitudes for the case of randomly distributed durations.

\begin{table}
\begin{ruledtabular}
\centering
 \begin{tabular}{c||c c|c c c } 
 \multirow{2}{*}{$P_{\tau}$} & \multirow{2}{*}{Estimated averages} & & \multicolumn{3}{c}{$\gamma$} \\
   &  & & $10^{-1}$ & 1 & 10\\ [0.5ex] 
 \hline
   \multirow{2}{*}{$\Ray$} & $\Sm{A}_{\est}/\Sm{A}$ & & \ldev{0.86} & \ldev{0.88} & \ldev{0.88}\\
  & $\Sm{w}_{\est}/\Sm{w}$ & & 0.99 & 1.03 & 1.00 \\[1ex]
  \hline
  \multirow{2}{*}{$\Gam(20)$} & $\Sm{A}_{\est}/\Sm{A}$ & & 0.96 & 0.97 & 1.06 \\
  & $\Sm{w}_{\est}/\Sm{w}$ & & 1.02 & 1.05 & \ldev{1.16} \\[1ex]
 \hline
  \multirow{2}{*}{$\Unif$} & $\Sm{A}_{\rm est}/\Sm{A} $ & & \ldev{0.80} & \ldev{0.83} & \ldev{0.80} \\
  & $\Sm{w}_{\est}/\Sm{w}$ & & 1.00 & 1.01 & 0.92\\[1ex]
 \end{tabular}
 \caption{Estimated averages of the amplitude and waiting times using different pulse duration distributions with the same average duration for all cases. \label{table:params-td-dist-amp-tw}}
 \end{ruledtabular}
\end{table}

The estimated average duration from fitting to the PSD is presented in \Tabref{tab:params-td-dist-est} for various duration distributions and intermittency parameters. The case of degenerately distributed pulse durations is equivalent to the case of exponentially distributed waiting times presented in \Tabref{tab:params-tw-dist-est}. Again, a $\Gam(20)$ distribution of durations does not appear to affect the results, but the average pulse duration is overestimated in both of the other cases. The case of uniformly distributed durations give the largest deviation in the estimated average pulse duration and will again be investigated in more detail.

\begin{figure}
    \centering
    \includegraphics[width=1.\textwidth]{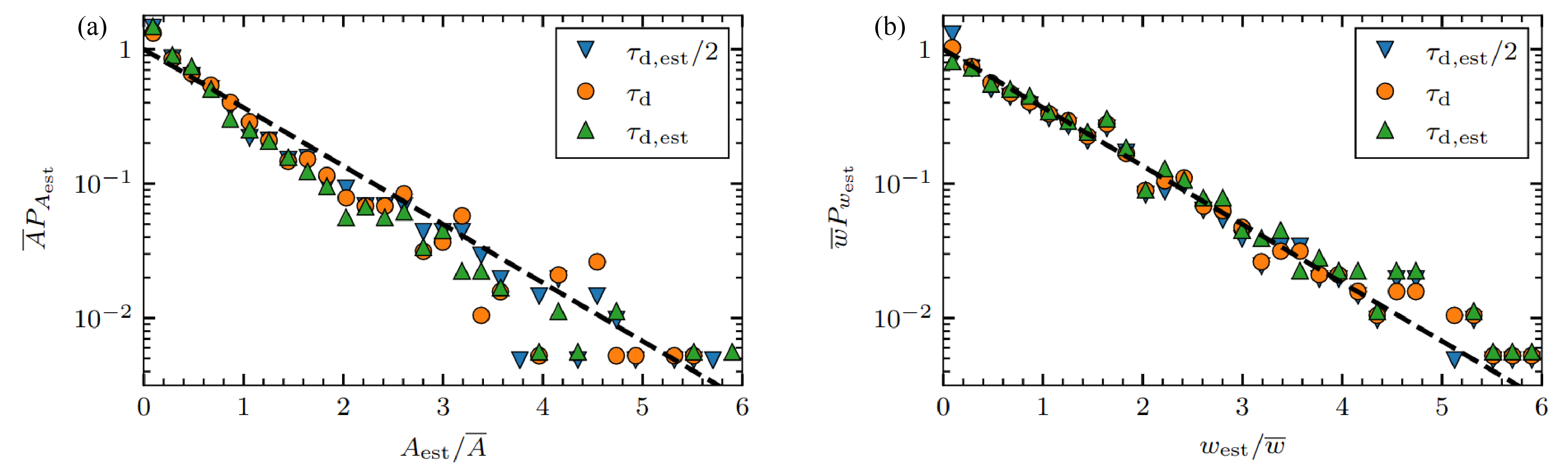}
    \caption{Estimated (a) amplitude distributions and (b) waiting time distributions for uniformly distributed duration times, $\tau\sim\Unif$ and $\gamma=10^{-1}$, using various assumed constant duration times. The black dashed lines are the references of the input distributions.}\label{fig:amp-tw-dist-tddist=unif}
\end{figure}

The estimated pulse amplitude and waiting time distributions for the case of uniformly distributed durations are presented in \Figref{fig:amp-tw-dist-tddist=unif}, and the estimated average amplitudes and waiting times are presented in \Tabref{tab:params-est-tddist-unif}. Again, we compare results from deconvolution with pulses using $\td$, $\tde$ and $\tde/2$ as pulse duration. Both the amplitude and waiting time distributions are well estimated. The estimated average waiting time is within 10\% of the original sample mean value for both estimated pulse durations used for deconvolution. However, the deconvolution with the reduced estimated pulse duration captures the average amplitude better than the full estimated duration time, consistent with the previous cases.


\begin{table}
\begin{ruledtabular}
\centering
 \begin{tabular}{c||c c c } 
\multirow{2}{*}{$\tau_{\mathrm{d, est}}/\Sm{\td}$} & \multicolumn{3}{c}{$\gamma$} \\
   &   $10^{-1}$ & 1  & 10 \\ [0.5ex] 
 \hline
  $\tau_\mathrm{d} \sim \Ray$ & \ldev{1.43} & \ldev{1.27} & \ldev{1.28} \\ 
  $\tau_\mathrm{d} \sim \Gam(20)$ & 1.02 & 1.10 & 0.98 \\ 
  $\tau_\mathrm{d} \sim \Unif$ & \ldev{1.54} & \ldev{1.45} & \ldev{1.39} \\ 
 \end{tabular}
\caption{Ratio between estimated pulse duration $\tau_{\mathrm{d, est}}$ and sample mean of the durations $\Sm{\tau}_\text{d}$ using different duration time distributions.}
\label{tab:params-td-dist-est}
\end{ruledtabular}
\end{table}

\begin{table}
\begin{ruledtabular}
\centering
 \begin{tabular}{c||c c c}
   Estimated averages &   $\tau_\mathrm{d}$ & $\tau_{\mathrm{d, est}}/2$  & $\tau_{\mathrm{d, est}}$ \\ [0.5ex] 
 \hline
  $\Sm{A}_{\est}/\Sm{A}$ & \ldev{0.80} & \ldev{0.80} & \ldev{0.72}\\ 
  $\Sm{w}_{\est}/\Sm{w}$ & 1.00 & 0.93 & 1.08 \\ 
 \end{tabular}
\caption{Table showing the rescaled estimated parameters from \Figref{fig:amp-tw-dist-tddist=unif} for uniformly distributed pulse durations using different average duration values in the deconvolution for $\gamma = 10^{-1}$. The corresponding estimated duration can be found in \Tabref{tab:params-td-dist-est}.}
\label{tab:params-est-tddist-unif}
\end{ruledtabular}
\end{table}

\subsection{Conclusion}

In this section, we have investigated the effect of estimating the pulse duration time from the PSD in the case of non-exponentially distributed waiting times, as well as for a distribution of pulse durations. If the shape of the power spectrum is similar to that of a single pulse (as it will be for uncorrelated pulses with an exponential waiting time distribution), the pulse duration is accurately estimated and there are no issues in applying the deconvolution method. Likewise, a narrow distribution of pulse durations ($\Gam(20)$ in our case) give reliable estimates of amplitudes and waiting times. However, broadly distributed pulse durations will lead to errors in the estimated averages, even if the average duration time is known exactly. It is again demonstrated that using a smaller duration time than the one estimated from the frequency spectrum is preferable, but did not improve on using the mean duration time.

\newpage
\section{Reconstruction of the pulse function}\label{sec:find-pulse-shape}
\begin{figure}
    \centering
    \includegraphics[width=0.48\textwidth]{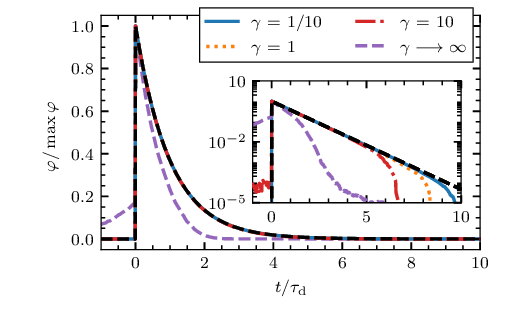}
    \caption{Reconstruction of a one-side exponential pulse function with different intermittency parameters without noise added. The inset shows the differences between the pulse reconstructions with semi-logarithmic axes. The black dashed line shows the true one-sided exponential pulse function.}
    \label{fig:phi-diff-gamma-no-noise}
\end{figure}

In some applications, the forcing is known or may be estimated, while the pulse function (or system response to the forcing) is unknown. It is clear from \Eqref{eq:sn-conv-disc} and the iteration scheme given by \Eqref{eq:deconv-scheme} that the particular interpretation of the vectors $f$ and $\snw$ does not affect the deconvolution algorithm if the known vector satisfies the conditions of non-negativity and a positive value at $t=0$. As such, we may consider $f$ a known forcing and $\snw$ an unknown pulse function and obtain the deconvolution algorithm by switching the symbols $f$ and $\snw$ in \Eqref{eq:deconv-scheme}:
\begin{equation} \label{eq:deconv-scheme-pulse}
    \snw^{(n+1)}_{j} = \snw^{(n)}_{j}\frac{\left(\Phi * \widehat{f}\right)_j + b}{\left(\snw^{(n)} * f * \widehat{f}\right)_j + b} .
\end{equation}
Here, we are interested in the direct result of the deconvolution, and so we do not expect the value of the intermittency parameter to significantly influence the result. This is confirmed by \Figref{fig:phi-diff-gamma-no-noise}, where we present reconstructions of the one-sided exponential pulse for different intermittency parameters from realizations of the model in the base case. In all cases with a finite intermittency parameter, the pulse reconstruction is reliable, although in the inset it is seen how a lager intermittency parameter pushes up the noise floor, thereby decreasing the time window of accurate reconstruction. Still, for $\gamma=10$, the effects are only seen two decades below the maximal pulse value. In \Figref{fig:phi-diff-gamma-no-noise}, we have also indicated the case $\gamma \to \infty$ by letting the forcing signal consist of independently and identically normally distributed random variables, with a constant added to make the forcing positive. In this most severe case, the pulse function is significantly affected, with a slight rise before the peak and correspondingly a faster decay after the peak. In the following, we will only consider the case of moderate pulse overlap, given by $\gamma=1$. For the results presented in this section, we use a portion of the synthetic time series and its known forcing, where both have a length of $2^{19}+1$ data points. The initial guess for estimating the pulse function is an array, also of size $2^{19}+1$ data points, containing a boxcar function centered at zero with an amplitude of one and a width of $2^{17} + 1$ data points, equivalent to about 1300 pulse duration times. All values outside the boxcar are set to zero. The boxcar is used to improve stability; allowing positive values over the entire estimated pulse function array can sometimes lead to spurious positive values at the far ends with a corresponding degradation of the pulse function in the center. Since zeros remain zero during the iteration of the algorithm, such effects are removed by the boxcar. Note that the boxcar is still huge compared to the expected size of the pulse it contains.

\begin{figure}
    \centering
    \includegraphics[width=0.48\textwidth]{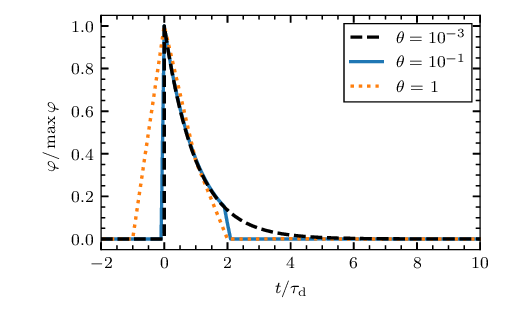}
    \caption{Reconstruction of a one-sided exponential pulse function (solid colored line and dotted colored line) for variously down-sampled signals with intermittency parameter $\gamma = 1$ and no noise added. The black dashed line shows the true one-sided exponential pulse.}
    \label{fig:phi-est-sampling}
\end{figure}

In \Figref{fig:phi-est-sampling}, reconstructed pulse functions from deconvolution of down-sampled model realizations using down-sampled forcing is presented. Here it is seen that the pulse function is not accurately reproduced if the process is undersampled. Note that for the comparison, the pulse functions have been rescaled to match the amplitude of the original pulse function. For $\theta = 10^{-1}$, the maximum value is $0.93$ and for $\theta = 1$, the maximum value is $0.62$. Thus, the reconstructed amplitude is affected as well as the pulse function. We note, however, that this result is sensitive to how undersampling affects the forcing, and in particular whether undersampling leads to losses of entire pulses or not. In the case presented here, loss of pulses in the forcing but not the signal is the major discrepancy between the original down-sampled forcing and the forcing estimated from the down-sampled signals.

\begin{figure}
  \centering
\includegraphics[width=1.\textwidth]{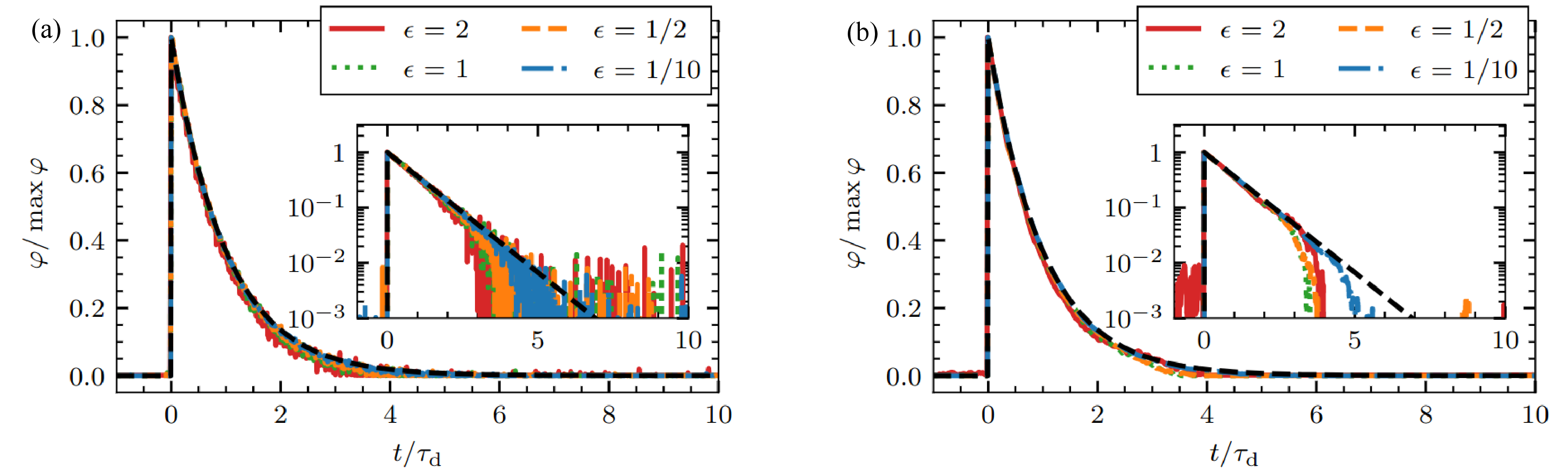}
  \caption{Reconstruction of a one-sided exponential pulse function using the modified RL-deconvolution for $\gamma=1$ and different values of the noise to signal ratio $\epsilon$. (a) Additive noise. (b) Dynamic noise. The black dashed lines represent the true one-sided exponential pulse function.}\label{fig:phi-recon-eps}
\end{figure}

We consider the effect of noise in \Figref{fig:phi-recon-eps}. Additive noise leads to noise in the tail, worse for higher $\epsilon$. Still, we reach noise rms two times the signal rms without significant deviations from the pulse function. Dynamical noise also distorts the pulse function, but not significantly, likely due to the noise in this case being convolved with the same pulse function as the forcing.

\begin{figure}
  \centering
  \includegraphics[width=0.48\textwidth]{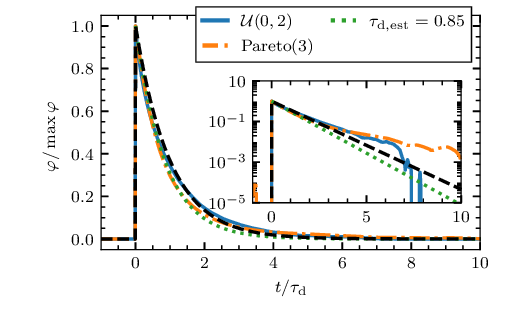}
  \caption{Pulse reconstruction with randomly distributed pulse durations for $\gamma = 1$ and no added noise. The inset shows the same with semi-logarithmic axes. The dotted line gives the best fit to the pulse function in the Pareto case. The black dashed line is the true one-sided exponential case.}\label{fig:phi-recon-td-dist}
\end{figure}

In \Figref{fig:phi-recon-td-dist}, we present reconstruction of the pulse function in the case of broad distribution of durations. The deconvolution method assumes all pulses have the same duration, but is shown to accurately reproduce the pulse function with largely the correct average duration. Narrower duration distributions ($\Ray$, $\Gam(20)$) were also attempted, but gave results indistinguishable from the true pulse function. Attempting to fit the result of the Pareto case to a single exponential on a linear scale gives $\tde = 0.85$, which is easily seen in the inset not to capture the correct pulse function. 

In conclusion, the deconvolution method can be used reliably in order to recover the pulse function from a given forcing. Only severe noise, undersampling or excessively broad pulse duration distributions lead to significant deviations from the average pulse function.

\newpage
\section{Discussion and conclusions}\label{sec:discussion-conclusion}

In this contribution, we have presented a novel method for extracting pulse amplitudes and arrival times from realizations of a stochastic process given by a superposition of pulses with fixed shape. The method relies on the ISRA deconvolution algorithm,which produces the maximum-likelihood solution to the deconvolution problem $\Phi = \varphi * f + X$ where $\Phi$ is a known signal, $\varphi$ is a known pulse or kernel function, $f$ is the forcing to be estimated and $X$ is normally distributed noise. Since the result of the deconvolution algorithm is the forcing time series $f$, a 3-point maxima is used to estimate pulse amplitudes and arrival times. 

For realizations of an intermittent process with high temporal resolution (sampling time $1/20$ times the average time between pulses or better and $1/10$ times the average pulse duration time or better), amplitude and waiting time distributions are well recovered in a variety of cases. Coarser sampling or more pulse overlap both lead to several pulses being counted as one, which distorts the estimated amplitude distribution and leads to overestimation of the average amplitude and waiting time. We note that this condition on the sampling time still allows for pulse overlap and pulses which are separated by two sampling times or more are robustly separated by the algorithm. Based on studies of numerous model realizations, it is recommend to use the approximate conditions $\gamma \theta = \dt / \Ttm{w} \leq 1/20$ and $\theta \leq 1/10$ to determine if the deconvolution will give reasonable estimates of mean values of pulse amplitudes and waiting times, and $\gamma \theta \leq 1/10$ if only the functional shape of the corresponding distributions are desired.

While the deconvolution method is only designed for positive valued signals and forcings, negative signal values may be accounted for by a straightforward modification of the algorithm. If negative values are present, we may recover both positive and negative parts by using the method on both the signal and its sign reversed version separately, and combine the results. While the method is not able to accurately resolve parts of data time series where fluctuations of different sign arrive close together, parts of the signal where one sign dominates are well reconstructed.

Noise may be handled, and relies at present on introducing an amplitude threshold in the 3-point maxima. The threshold performs well recovering the tail of the amplitude and waiting time distributions. Average pulse amplitudes are estimated to within 15\% of their true value, while mean waiting times are estimated to within a factor two.

If the pulse duration is not known before applying the deconvolution method, it may be estimated from the frequency PSD of the process. The spectrum is insensitive to pulse overlap and amplitude distribution and robust to non-exponential waiting time distributions as well as distributions of the pulse duration. It is demonstrated that for intermittent processes, underestimation of the pulse duration has little to no effect on the estimation of pulse amplitudes and arrivals while overestimation of the pulse duration has significant implications. Broad distributions in waiting times or durations lead to overestimation of the average duration, which in turn distorts average amplitude and waiting time estimated from the deconvolution. It is shown that in these cases, performing the deconvolution with a pulse duration half the estimated value from the PSD improves the results.

Lastly, if the forcing is known but the pulse function is unknown, the ISRA algorithm may be employed straightforwardly. We have demonstrated that the reproduction of the pulse function is excellent for all but the most severe undersampling or noise. Even if there is a narrow distribution of the pulse durations, the algorithm recovers the pulse function with the average duration time.

For real data, we advise a procedure as in Refs.~\onlinecite{kube-2020} and \onlinecite{theodorsen-pop-2018}. First, the analysis leading to estimates of the stochastic model parameters is performed. Then, the deconvolution is performed on the empirical data. Finally, synthetic data from model realizations with the estimated model parameters is made and analysed in the same manner as the empirical data. Results from analysis of the synthetic data should then be compared to the results of the measurement time series. Ideally, this should be carried out as a Monte-Carlo study with multiple model realizations, demonstrating that the results of the deconvolution are within the expected errors for synthetic data.

In conclusion, the deconvolution algorithm is shown to recover amplitude and waiting time distributions from realizations of an intermittent process even in the presence of significant pulse overlap, addition of noise, and deviations from the expected pulse function. For all signals considered in this contribution, the underlying ISRA method recovers the forcing admirably and only the basic information loss associated with the finite sampling of a continuous signal affects the reconstruction of amplitude and waiting time distributions.


\begin{acknowledgments}
This work was supported by Tromsø Research Foundation under grant number 19\_SG\_AT and the UiT Aurora Centre Program, UiT The Arctic University of Norway (2020).
\end{acknowledgments}
\appendix

\section{Effect of sampling on event recollection}\label{app:number-events}

In this section, we investigate the information loss associated with poor sampling of the filtered Poisson process (FPP). Consider a Poisson point process $K(T)$ on the interval $[0, T)$ with rate parameter $1/\gamma$. For a given realization, the arrivals of $K$ events are uniformly distributed on the interval. The interval is discretized into $N$ time steps of size $\triangle_t=T/N$. For reference, we note that
\begin{equation}\label{eq:mean-K}
    \Tm{K} = \gamma \theta N ,
\end{equation}
where $\theta=\triangle_t/\td$. We only record if events occur in a given time step, but not how many events occur. We therefore move from the process $K(T)$ to the process $F(N)$, denoting locations with events. By necessity, $F \leq K$ and $F \leq N$. For each of the $N$ time steps, the probability of receiving events is $1-\text{Pr}$[No events in time $\theta$], which from the Poisson distribution of $K$ is $1-\exp(-\gamma \theta)$. Therefore, the probability mass function (PMF) of $F$ is a Binomial distribution over $N$ trials with success probability $1-\exp(-\gamma \theta)$,
\begin{equation}\label{eq:pmf-f-gt}
 P_F(f ; \gamma, \theta, N) = \binom{N}{f} \exp(-\gamma \theta)^{N-f} [1-\exp(-\gamma \theta)]^f.   
\end{equation}
The mean value is given by
\begin{equation}\label{eq:mean-F}
    \Tm{F} = N \left[ 1 - \exp(-\gamma \theta) \right],
\end{equation}
For $\gamma \theta \ll 1$, the exponential in \Eqref{eq:mean-F} can be expanded and $\Ttm{F} \approx \Ttm{K}$. However, $\gamma \approx 1/\theta$ gives $\Ttm{K} \approx N$ but $\Ttm{F} \approx 0.6 N$ as many events arrive at the same discrete time location.

Let us now investigate the effect of the 3-point maxima peak finding algorithm. Letting $M$ denote the number of maxima, we do the following approximation (here ``cluster of size $k$" means $k$ consecutive filled time steps with empty time steps at each end)
\begin{eqnarray}\label{eq:approx-mean-maxima-start}
    M = \sum_c \sum_m ( {\rm number\, of\, maxima\, in\, cluster\, } m {\rm \, of\, size\, }c) \nonumber\\
    \Tm{M} \approx N \sum_c P[{\rm cluster\, of\, size\, }c]\Tm{{\rm  number\, of\, maxima\, in\, clusters\, of\, size\, }c} 
\end{eqnarray}
Letting $C$ be the cluster size, the probability of having a cluster of size $c$ is given by
\begin{equation}\label{eq:cluster-size}
    \text{Pr}[C=c] = \left( 1 - e^{-\gamma \theta} \right)^c \left( e^{-\gamma \theta} \right)^2.
\end{equation}
To find the number of maxima per cluster, we argue as follows: As the amplitudes are independently and identically distributed, so are the values of the forcing at neighboring time steps. As such, all permutations of the forcing values in a cluster are equally likely. For $c=1$ and $c=2$, there is obviously just one maximum.  For $c=3$, there are 6 permutations, two of which give two maxima and the rest give one, for an average of $4/3$ maxima in the cluster. By going from two to three data points, we had a $1/3$ chance of adding an extra maximum. Adding further data points to the end of the sequence each time gives an additional $1/3$ chance of a new maxima, so the average number of maxima should increase by $1/3$ per new data point. Aided by brute force investigation of all ordered sequences up to $c=10$, we guess that the average number of maxima in an ordered sequence of size $c \geq 2$ is $(c+1)/3$, including the end points. Adding $1/3$ to this gives $((c+1)+1)/3$, the average number of maxima in a sequence of $c+1$ data points. We therefore have that the approximation in \Eqref{eq:approx-mean-maxima-start} can be written as
\begin{eqnarray}\label{eq:approx-mean-M}
    \frac{\Tm{M}}{N} \approx \text{Pr}[C=1] + \sum\limits_{c=2}^\infty \text{Pr}[C=c] \frac{c+1}{3} \nonumber \\
    \frac{\Tm{M}}{N} \approx \frac{1 - e^{-\gamma \theta}}{3} \left[ 1 + e^{-\gamma \theta} + \left( e^{-\gamma \theta} \right)^2 \right]
\end{eqnarray}
where we have used $\text{Pr}[C=c]$ from \Eqref{eq:cluster-size}. This expression fits our expectations: 
For very small $\gamma \theta$, we mainly expect clusters of size $1$, and \Eqref{eq:approx-mean-M} approaches \Eqref{eq:mean-F}. For very large $\gamma \theta$, the entire time series is likely to be filled, so we expect about $N/3$ maxima following the discussion above. 

\begin{figure}
  \centering
  \includegraphics[width=0.48\textwidth]{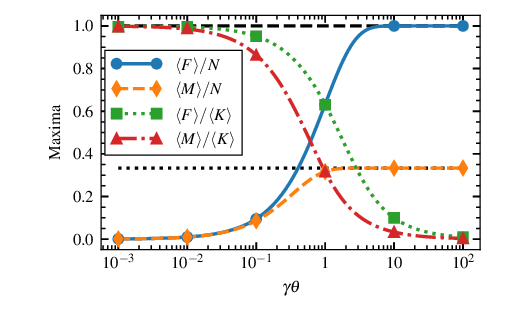}
    \caption{\label{fig:comparison-K-F-M} Comparison of the expected number of events $\Tm{K}$, the expected number of time steps with events $\Tm{F}$ and the expected number of events found after a 3-point maxima $\Tm{M}$ as a function of $\gamma \theta$. In all cases, the number of data points was $N=10^5$ and the normalized time step was $\theta=10^{-2}$. The lines give analytical approximations, and the symbols are results from numerical simulations.}
\end{figure}

In \Figref{fig:comparison-K-F-M}, we compare the analytical results of this section with results from a Monte-Carlo study. The lines give the analytical predictions, while the points give mean values of $20$ realizations of the base case with $N=10^5$ and $\theta=10^{-2}$ for various $\gamma$. One standard deviation would give error bars smaller than the plot symbols. \Eqref{eq:approx-mean-M} is an excellent approximation to the average number of maxima for large $N$, and that the full distribution of the number of events is narrow, justifying the use of $\Ttm{M}$ instead of the full distribution. 

For the deconvolution method, $\Ttm{M}/\Ttm{K}$ is the most significant number, as this gives a measure of how well we may hope to recreate the individual pulses in the time series. The lower this ratio is, the more separate pulses are counted as one. If we had a way of avoiding the 3-point maxima, we would still have the coarse-graining introduced by discretizing the time series. In this ideal case, $\Ttm{F}/\Ttm{K}$ is the important ratio.

\section{Definition of distributions}\label{app:dist-def}
In this section, we list the distributions used in this manuscript. In all cases, we give the distributions in terms of their mean value $\Ttm{X} = \mu$, so that for any realization $\{X_k\}_{k=1}^K$ listed here, we may get amplitudes, waiting times or pulse durations by setting $\mu = \Ttm{A}$, $\mu = \tw $, or $\mu = \td$ respectively. Unless otherwise noted, all distributions used are positive definite, so we only give the distributions for $x>0$. For $x<0$, we have $P_X(x)=0$.

\paragraph{The exponential distribution} is given by
\begin{equation}\label{eq:exp-dist-def}
    P_X(x) = \frac{1}{\mu} \exp\left( - \frac{x}{\mu} \right).
\end{equation}
We denote an exponentially distributed random variable by $X \sim {\rm Exp}$

\paragraph{The gamma distribution} is given by
\begin{equation}\label{eq:gamma-dist-def}
    P_X(x;k) = \frac{k^k}{\Gamma(k) \mu^k} x^{k-1} \exp\left(- \frac{k x}{\mu}\right).
\end{equation}
This distribution has one free parameter, the shape parameter $k$. For $k=1$, this coincides with the exponential distribution. In the main text, the gamma distribution is denoted by $X \sim \Gam(k)$

\paragraph{The Rayleigh distribution} is given by
\begin{equation}\label{eq:ray-dist-def}
    P_X(x) = \frac{\pi x}{2 \mu^2} \exp\left(-\frac{\pi x^2}{4 \mu^2} \right).
\end{equation}
We denote a Rayleigh distributed random variable by $X \sim \Ray$

\paragraph{The Pareto distribution} is given by 
\begin{equation}\label{eq:pareto-dist-def}
    P_X(x) = \frac{(\alpha-2)^{\alpha-1}}{(\alpha-1)^{\alpha-2}} \frac{\mu^{\alpha-1}}{x^\alpha}, \quad x \geq \frac{\alpha-2}{\alpha-1} \mu.
\end{equation}
As we require a well-defined mean for all random variables, we demand $\alpha>2$.
We denote a Pareto distributed random variable by $X \sim \Par(\alpha)$. Note that following this definition, the PDF decays as $x^{-\alpha}$, while in the standard definition, it is the cumulative distribution function which decays as $x^{-\alpha}$.

\paragraph{The degenerate distribution} is given by
\begin{equation}\label{eq:deg-dist-def}
    P_X(x) = \delta(x-\mu).
\end{equation}
We denote a degenerately distributed random variable by $X \sim \Deg$.

\paragraph{The uniform distribution} is given by
\begin{equation}\label{eq:unif-dist-def}
    P_X(x) = \begin{cases}
    \frac{1}{2 \mu}, & 0\leq x \leq 2\mu \\
    0, & \text{else}.
    \end{cases}
\end{equation}
Note that this is the broadest possible non-negative uniform distribution with mean $\mu$. We denote the uniform distribution by $X \sim \Unif$.

\section{Pearson correlation coefficients}\label{app:pearson}
In this appendix, we present Pearson correlation coefficients between the true forcing $f$ and the forcing $f_\est$ from estimated amplitudes and arrival times. These results are presented for completion, in general the conclusions that may be drawn from these tables are the same as may be drawn from the distributions and mean values in the main text. To guide the eye and highlight this correspondence, we have marked correlations below $0.9$ in italics.

Comparing \Tabsref{table:eps-pearson-no-thresh} (for noisy signals without thresholding) and \ref{table:eps-pearson} (for noisy signals with thresholding), we see that the thresholding, although it improves the estimate of the amplitude- and waiting time distributions, uniformly lead to larger differences between $f$ and $f_\est$. If the thresholding were able to separate between events and noise, we would expect the thresholding to improve the correlation. From the Tables, it is evident that it removes many true events as well.

 \begin{table}[h!]
 \begin{ruledtabular}
 \centering
  \begin{tabular}{c||c c c c c} 
   \multirow{2}{*}{}  & \multicolumn{5}{c}{$\gamma$} \\
   & 1/10 & 1  & 5 & 10 & 50 \\ [0.5ex] 
  \hline
   $A \sim \Exp$ & 0.95 & 0.93 & 0.92 & 0.91 & \ldev{0.86}\\ 
     $w \sim \Exp$ & 0.95 & 0.93 & 0.92 & 0.91 & \ldev{0.86}\\ 
  \end{tabular}
 \caption{Table of the Pearson correlation coefficients between the true forcing and the forcing from estimated amplitudes and arrivals corresponding to the the base case, \Figref{fig:exp-amp-tw-dist} and \Tabref{table:est-mean-exp}.}
 \label{tab:pearson-exp-amp-tw-dist}
 \end{ruledtabular}
 \end{table}

\begin{table}[h!]
\begin{ruledtabular}
\centering
 \begin{tabular}{c||c c c c c} 
    & \multicolumn{5}{c}{$\gamma$} \\
   & $10^{-1}$ & 1  & 5 & 10 & 50 \\ [0.5ex] 
 \hline
  $\Ray$ & 0.94 & 0.92 & 0.91 & 0.90 & \ldev{0.82}\\ 
  $\mathrm{Pareto(3)}$ & 0.92 & 0.93 & 0.96 & 0.92 & 0.90 \\[1ex]
  $\Unif$ &  0.94& 0.92 & 0.91 & 0.90 & \ldev{0.82} \\[1ex]
  $\Deg$ & 0.93 & 0.92 & 0.90 & \ldev{0.88} & \ldev{0.77} \\[1ex]
 \end{tabular}
\caption{Pearson correlation coefficients between the true forcing and the forcing from estimated amplitudes and arrivals for different amplitude distributions, corresponding to \Figref{fig:amp-dist-var} and \Tabref{table:amp-est-mean}.}
\label{tab:pearson-diff-amp-dist}
\end{ruledtabular}
\end{table}

\begin{table}[h!]
\begin{ruledtabular}
\centering
 \begin{tabular}{c||c c c c c} 
    & \multicolumn{5}{c}{$\gamma$} \\
 & $10^{-1}$ & 1  & 5 & 10 & 50 \\ [0.5ex] 
 \hline
  $\Ray$ & 0.90 & 0.93 & 0.92 & 0.92 & \ldev{0.87}\\ 
  $\mathrm{Pareto(3)}$ & 0.92 & 0.93 & 0.92 & 0.92 & 0.91\\ [1ex]
    $\Unif$ & 0.91 & 0.93 & 0.91 & 0.91 & \ldev{0.87} \\[1ex]
  $\Deg$ & 1.00 & 1.00 & 0.93 & 0.93 & 0.90 \\[1ex]
 \end{tabular}
\caption{Pearson correlation coefficients between the true forcing and the forcing from estimated amplitudes and arrivals for different waiting time distributions, corresponding to \Figref{fig:tw-dist-var} and \Tabref{table:tw-est-mean}}
\label{tab:pearson-diff-tw-dist}
\end{ruledtabular}
\end{table}

\begin{table}
\begin{ruledtabular}
\centering
 \begin{tabular}{c||c c|c c c} 
 & \multirow{2}{*}{$\epsilon$} & & \multicolumn{3}{c}{$\gamma$} \\
   &  & & $10^{-1}$  & 1 & 10 \\ [0.5ex] 
 \hline
  \multirow{3}{*}{Additive} & $10^{-1}$ & & 0.99 & 0.95 & \ldev{0.74}\\ 
   & 1/2 & & 0.97 & \ldev{0.83} & \ldev{0.43}\\
   & 1 & & 0.95 & \ldev{0.73} & \ldev{0.30}  \\[1ex]
 \hline
  \multirow{3}{*}{Dynamic} & $10^{-1}$ & & 1.00 & 0.99 & 0.97\\ 
   & 1/2 & & 0.98 & 0.93 & \ldev{0.88}\\
   & 1 & & 0.96 & \ldev{0.88} & \ldev{0.78}   \\[1ex]

 \end{tabular}
  \caption{Table showing the values of the Pearson correlation coefficients between the true forcing and the estimate forcing for different noise to signal ratios, corresponding to \Figref{fig:amp-tw-noise-no-thresh}. \label{table:eps-pearson-no-thresh}}
 \end{ruledtabular}
\end{table}

\begin{table}[h!]
\begin{ruledtabular}
\centering
 \begin{tabular}{c||c c|c c c} 
  & \multirow{2}{*}{$\epsilon$} & & \multicolumn{3}{c}{$\gamma$} \\
   &  & & $10^{-1}$  & 1 & 10 \\ [0.5ex] 
 \hline
  \multirow{3}{*}{Additive} & $10^{-1}$ & & 0.95 & \ldev{0.89} & \ldev{0.67}\\ 
   & 1/2 & & 0.93 & \ldev{0.77} & \ldev{0.37}\\
   & 1 & & 0.92 & \ldev{0.68} & \ldev{0.24}  \\[1ex]
 \hline
  \multirow{3}{*}{Dynamic} & $10^{-1}$ & & 0.95 & 0.91 & \ldev{0.85} \\ 
  & 1/2 & & 0.94 & \ldev{0.85} & \ldev{0.65}  \\
  & 1 & & 0.93 & \ldev{0.78} & \ldev{0.49} \\[1ex]
 \end{tabular}
 \caption{Pearson correlation coefficients between the true forcing and the forcing using estimated amplitudes and arrival times with thresholding, for various intermittency parameters and noise to signal ratios corresponding to \Figsref{fig:amp-noise} and \ref{fig:twdist-noise} and \Tabsref{table:amp-est-var-eps-gamma} and \ref{table:tw-est-var-eps-gamma-rescaled}.  \label{table:eps-pearson}}
 \end{ruledtabular}
\end{table}

\begin{table}[h!]
\begin{ruledtabular}
\centering
 \begin{tabular}{c||c c c c c} 
\multirow{2}{*}{$\gamma$} & \multicolumn{5}{c}{$\tda/\td$} \\
   &   $10^{-1}$ & 1/2 & 1  & 2 & 10 \\ [0.5ex] 
 \hline
   $10^{-1}$ & 0.95 & 0.95 & 0.95 & 0.94 & \ldev{0.77}\\ 
  1 & 0.92 & 0.93 & 0.93 & \ldev{0.88} & \ldev{0.61}\\ 
  10 & \ldev{0.87} & 0.91 & 0.91 & \ldev{0.84} & \ldev{0.53}\\
 \end{tabular}
\caption{Pearson correlation coefficients between the true forcing and the estimated amplitudes and arrival times using 3-point maxima for various pulse duration times and intermittency values. This table corresponds to \Figref{fig:td-wrong-recon} and \Tabsref{tab:mean_est_tw_diff_td} and \ref{tab:mean-est-amp-diff-td}.}
\label{tab:diff_td_pearson}
\end{ruledtabular}
\end{table}

\clearpage
\bibliographystyle{apsrev4-2}
\bibliography{sources}

\end{document}